\documentclass[twocolumn]{aastex701}

\newcommand{\kms}{km\,s$^{-1}$}
\newcommand{\hh}{\ensuremath{\mathrm{H}_2}}
\newcommand{\ts}{$\mathrm{\tau_{Si}}$}
\newcommand{\xitr}{\ensuremath{\xi(T_{\mathrm{rot}})}}

\received{April 9, 2026}
\revised{May 15, 2026}
\accepted{May 16, 2026}

\usepackage{amsmath}
\usepackage{siunitx}

\begin{document}

\title{JWST Observations of Starbursts: Molecular Hydrogen Excitation and Disequilibrium in M82}

\author[0000-0003-0014-0508]{Sara E. Duval}
\affiliation{Ritter Astrophysical Research Center, Department of Physics \& Astronomy, University of Toledo, 2801 West Bancroft Street, Toledo, OH 43606, USA}
\email[show]{saraduval018@gmail.com}

\author[0000-0003-1545-5078]{J.D.T. Smith}
\affiliation{Ritter Astrophysical Research Center, Department of Physics \& Astronomy, University of Toledo, 2801 West Bancroft Street, Toledo, OH 43606, USA}
\email{jd.smith@utoledo.edu}

\author[0000-0002-5480-5686]{Alberto D. Bolatto}
\affiliation{Department of Astronomy, University of Maryland, College Park, MD 20742, USA}
\affiliation{Joint Space-Science Institute, University of Maryland, College Park, MD 20742, USA}
\email{bolatto@umd.edu}

\author[0000-0002-0846-936X]{B.T. Draine}
\affiliation{Princeton University, Princeton, NJ 08540, USA}
\email{draine@astro.princeton.edu}

\author[0000-0001-8490-6632]{Thomas S.-Y. Lai}
\affiliation{IPAC, California Institute of Technology, 1200 East California Boulevard, Pasadena, CA 91125, USA}
\email{shaoyu@ipac.caltech.edu}

\author[0000-0002-4378-8534]{Karin M. Sandstrom}
\affiliation{Department of Astronomy \& Astrophysics, University of California, San Diego, La Jolla, CA 92093, USA}
\email{kmsandstrom@ucsd.edu}

\author[0000-0001-6708-1317]{Simon C.~O.\ Glover}
\affiliation{Universit\"{a}t Heidelberg, Zentrum f\"{u}r Astronomie, Institut f\"{u}r Theoretische Astrophysik, Albert-Ueberle-Str.\ 2, 69120 Heidelberg, Germany}
\email{glover@uni-heidelberg.de}

\author[0000-0002-0560-3172]{Ralf S.\ Klessen}
\affiliation{Universit\"{a}t Heidelberg, Zentrum f\"{u}r Astronomie, Institut f\"{u}r Theoretische Astrophysik, Albert-Ueberle-Str.\ 2, 69120 Heidelberg, Germany}
\affiliation{Universit\"{a}t Heidelberg, Interdisziplin\"{a}res Zentrum f\"{u}r Wissenschaftliches Rechnen, Im Neuenheimer Feld 225, 69120 Heidelberg, Germany}
\email{klessen@uni-heidelberg.de}

\author[0000-0001-8782-1992]{Elisabeth A.C. Mills}
\affiliation{Department of Physics and Astronomy, University of Kansas, 1251 Wescoe Hall Drive, Lawrence, KS 66045, USA}
\email{eacmills@ku.edu}

\author[0000-0003-2508-2586]{Rebecca C. Levy}
\affiliation{Space Telescope Science Institute, 3700 San Martin Drive, Baltimore, MD 21218, USA}
\email{rlevy.astro@gmail.com}

\author[0000-0002-3158-6820]{Sylvain Veilleux}
\affiliation{Department of Astronomy, University of Maryland, College Park, MD 20742, USA}
\affiliation{Joint Space-Science Institute, University of Maryland, College Park, MD 20742, USA}
\email{veilleux@umd.edu}

\author[0000-0002-5782-9093]{Daniel~A.~Dale}
\affiliation{Department of Physics and Astronomy, University of Wyoming, Laramie, WY 82071, USA}
\email{ddale@uwyo.edu}

\author[0000-0001-5042-3421]{Aditya Togi}
\affiliation{Department of Physics, 601 University Dr., Texas State University, San Marcos, TX 78666, USA}
\email{togiaditya@gmail.com}

\author[0000-0001-5434-5942]{Paul P. van der Werf}
\affiliation{Leiden Observatory, Leiden University, P.O.~Box 9513, 2300 RA Leiden, The Netherlands}
\email{pvdwerf@strw.leidenuniv.nl}

\author[0000-0002-5877-379X]{Vicente Villanueva}
\affiliation{Instituto de Estudios Astrofísicos, Facultad de Ingeniería y Ciencias, Universidad Diego Portales, Av. Ejército Libertador 441, 8370191 Santiago, Chile}
\affiliation{Millennium Nucleus for Galaxies, MINGAL}
\email{vicente.avl365@gmail.com}

\author[0000-0003-3633-0098,sname='Siwakoti']{Utsav Siwakoti}
\affiliation{Department of Physics and Astronomy, University of Kansas, 1251 Wescoe Hall Drive, Lawrence, KS 66045, USA}
\email{siwakotiutsav@ku.edu}

\author[0000-0002-9511-1330]{Serena A. Cronin}
\affiliation{Department of Astronomy, University of Maryland, College Park, MD 20742, USA}
\email{cronin@umd.edu}

\author[0000-0003-0605-8732]{Evan D.\ Skillman}
\email{skill001@umn.edu}
\affiliation{Minnesota Institute for Astrophysics, University of Minnesota, Minneapolis, MN 55455, USA}

\author[0000-0003-0645-5260]{Deanne B. Fisher}
\affiliation{Centre for Astrophysics and Supercomputing, Swinburne University of Technology, Hawthorn, VIC 3122, Australia}
\email{dfisher@swin.edu.au}

\author[0000-0003-4209-1599]{Yu-Hsuan Teng}
\affiliation{Department of Astronomy, University of Maryland, 4296 Stadium Drive, College Park, MD 20742, USA}
\email{yhteng@umd.edu}

\author[0000-0001-9436-9471]{David S. Meier}
\affiliation{New Mexico Institute of Mining and Technology, 801 Leroy Place, Socorro, NM, 87801, USA}
\email{david.meier@nmt.edu}

\author[0000-0002-3952-8588]{Leindert A. Boogaard} \affiliation{Leiden Observatory, Leiden University, PO Box 9513, NL-2300 RA Leiden, The Netherlands}
\email{boogaard@strw.leidenuniv.nl}

\author[0000-0003-1356-1096]{Elizabeth Tarantino}
\affiliation{Space Telescope Science Institute, 3700 San Martin Drive, Baltimore, MD 21218, USA}
\email[hide]{etarantino@stsci.edu}  

\author[0000-0003-4023-8657]{Laura Lenki\'{c}}
\affiliation{IPAC, California Institute of Technology, 1200 East California Boulevard, Pasadena, CA 91125, USA}
\email{llenkic@ipac.caltech.edu}

\author[0000-0002-2775-0595]{Rodrigo Herrera-Camus}
\affiliation{Departamento de Astronomía, Universidad de Concepción, Barrio Universitario, Concepción, Chile}
\affiliation{Millennium Nucleus for Galaxies, MINGAL}
\email{rhc@astro-udec.cl}

\author[0000-0003-4793-7880]{Fabian Walter}
\affiliation{Max Planck Institut f\"ur Astronomie, K\"onigstuhl 17, D-69117 Heidelberg, Germany}
\email{walter@mpia.de}

\author[0009-0001-6844-9758]{Patricia~A.~Arens}
\affiliation{Centre for Astrophysics and Supercomputing, Swinburne University of Technology, Hawthorn, VIC 3122, Australia}
\email{patricia.a.arens@gmail.com}

\begin{abstract}
% CONTEXT
Emission from the pure rotational transitions of \hh\ traces warm molecular gas, providing insight into its temperature distribution and local heating conditions.
% AIMS
We have extended previous power-law \hh\ temperature models to account for differential extinction by dust as well as non-equilibrium ortho-to-para-\hh\ ratios (OPR). 
% METHODS
The turbulent environment of the M82 starburst offers a unique opportunity to study \hh\ out of equilibrium conditions, using $\sim$15 pc spatially resolved measurements from MIRI/MRS on JWST.
With extensive detections of \hh\ S(1)--S(7), we use our model to assess spatial variations in local heating conditions of molecular gas across a $\sim\,$500~pc region of the M82 central starburst.
% RESULTS
The average slope of the recovered \hh\ power law temperature distribution is consistent with prior studies, and the slope strongly anti-correlates with relative [\ion{Fe}{2}]/\hh\ S(1)--S(2) strength, pointing to the importance of shock-heating.
Our models indicate that the OPR is, on average, about half of its equilibrium value. This suppression is attributed to cloud mixing timescales which are short compared to timescales for spin conversion, with molecular gas remembering its ``cooler past''.  
% CONCLUSIONS
By accounting for OPR disequilibrium, we can identify instances of recent and rapid heating to better understand the flow of energy through the interstellar medium and track its thermal history.
\end{abstract}


\section{Introduction} 
\label{sec:intro}
By mass and number, molecular hydrogen (\hh) dominates the molecular composition of the universe. Despite lacking a dipole moment, its quadrupole transitions can be observed in the near-infrared (NIR) and mid-infrared (MIR) portions of the electromagnetic spectrum. The pure rotational transitions of H$_2$ detectable with MIRI/MRS, $J$=3--1 up to $J$=10--8 (S(1)--S(8)), probe gas temperatures of $\sim$\,100--1000\,K and emit in the MIR between $\sim$\,5--30\,µm \citep{Rigopoulou2002, Roussel2007}.  

The James Webb Space Telescope (JWST) offers unprecedented sensitivity for studying \hh\ with the Mid-infrared Instrument Medium Resolution Spectrometer (MIRI/MRS). With spectral resolution R$\sim$1300--3700 \citep{Wells2015}, JWST/MIRI MRS provides higher resolution measurements than the Infrared Spectrograph on board \textit{Spitzer} (R$\lesssim$600; \citealt{Houck2004}). Obtaining spatially resolved spectroscopy, MIRI/MRS measures pure rotational transitions S(1)--S(7) with a level of detail that previously was not possible. Measuring the \hh\ emission spectrum is important because of the information that H$_2$ excitation provides about local heating conditions of the gas and the ways that these conditions correlate with physical events such as winds and outflows from star formation and supernovae \citep{Heckman1990}. 

Previous work on \hh\ excitation in galaxies revealed information about the temperature distribution of the molecular gas based on the observed column densities as measured by \textit{Spitzer} and the Infrared Space Observatory. Typically this has been done by modeling the temperature distribution with either 1- or 2-component temperature fits \citep{Fuente1999, Neufeld2006, Roussel2007, PW2023}, or a power-law distribution \citep{TS2016, Zakamska2010}. Most models assume an equilibrium ortho-to-para \hh\ ratio. Beyond the temperature distribution of the gas, the abundance ratio of ortho-to-para-\hh\ encodes information about its heating history. In this work, we expand power-law models to capture information on the equilibrium conditions of the gas, which is of particular relevance for gas which has been cooled or heated rapidly on timescales shorter than the inter-species conversion timescale. 

A particularly interesting laboratory for studying H$_2$ excitation is the nearby edge-on starburst galaxy, M82 ($d$\,=\,3.6 Mpc; \citealt{Freedman1994}). The central $\sim$ 1~kpc of M82 has undergone two bursts of star formation occurring 8--15\,Myr ago in the central 500\,pc and 4--6\,Myr ago in a circumnuclear ring and the stellar bar \citep{FS2003}, which has decreased in the last 5\,Myr due to negative feedback processes \citep{Beirao2008}. The successive bursts of star formation produce FUV radiation and shock waves which provide both thermal and non-thermal excitation mechanisms for H$_2$ \citep{Heckman1990, Lord1996}. \citet{Bolatto2024} and \citet{Fisher2025} present imaging of the M82 nuclear starburst and the base of its prominent galactic wind with the JWST near-Infrared camera instrument, revealing filaments and plumes of 3.3\,µm polycyclic aromatic hydrocarbon (PAH) emission outflowing from the base of the galactic wind. In this paper, we leverage resolved \hh\ spectroscopy across the central $\sim$ 500~pc of M82 to assess the conditions and thermal history of its warm molecular gas content.

This work is organized as follows: In  Section~\ref{sec:bkg}, we discuss background information relevant to the exploration of H$_2$ excitation; in Section~\ref{sec:data} we detail the steps taken to reduce the MIRI/MRS data for M82 and to perform our spectral extractions; in Section~\ref{sec:ksimodel} we introduce our model of \hh\ excitation with subsequent results and discussion in Section~\ref{sec:results} and Section~\ref{sec:disc}, respectively; Section~\ref{sec:shortcomings} addresses caveats of the model and Section~\ref{sec:conc} presents our conclusions.

\section{\hh\ Excitation and Ortho-to-Para Ratio}
\label{sec:bkg}
Emission from the pure rotational transitions of \hh\ probes the molecular gas reservoirs in galaxies. Understanding the mechanisms that excite the \hh\ molecule and the reactions that convert its nuclear spin between its two states, ortho- and para-\hh, reveal information about local heating conditions of the gas.

\subsection{Excitation Mechanisms}
\label{sec:exc_mech}
\hh\ may be excited by both thermal and non-thermal mechanisms. Thermally, dynamic activity such as shocks, molecular outflows, and supernova remnants can promote collisions between hydrogen molecules and atoms, other molecules, and fast electrons \citep{BlackDalgarno1976, ShullBeckwith1982, HM1989, Rosenthal2000}. Non-thermal excitation processes include the absorption of UV-photons by H$_2$ where fluorescence leads to the population of excited rovibrational levels of the ground electronic state \citep{BlackDalgarno1976, BlackVanDishoeck1987}. Stars in photo-dissociation regions (PDRs) provide far-UV (FUV) photons to radiatively excite \hh\ \citep{Davies2003}.

\subsection{Equilibrium Ortho-to-Para Ratio}
\label{sec:equilOPR}
H$_2$ effectively behaves as two species, ortho-H$_2$ and para-H$_2$, differentiated by the parallel and anti-parallel alignment of their nuclear spins, respectively. Their individual contributions to total H$_2$ gas content are revealed through the ortho-to-para ratio (OPR). In equilibrium, the OPR$_{\mathrm{eq}}$ is given by the ratio of their respective partition functions, 

\begin{equation}
    {\rm OPR}_{\mathrm{eq}} = \frac{Z_o(T_{\mathrm{rot}})}{Z_p(T_\mathrm{{rot}})}
    \label{eq:OPR_eq}
\end{equation}
where $T_\mathrm{{rot}}$ is the rotational temperature of the molecular gas and 
\begin{equation}
    Z_{o}(T_\mathrm{{rot}}) = \sum_{J=\mathrm{odd}}^{\infty} g_J e ^{\frac{-E_J}{kT_\mathrm{{rot}}}}
    \label{eq:partitionO}
\end{equation}

\begin{equation}
    Z_{p}(T_{\mathrm{rot}}) = \sum_{J=\mathrm{even}}^{\infty} g_J e^{\frac{-E_J}{kT_{\mathrm{rot}}}}
    \label{eq:partitionP}
\end{equation}
where $J$ is the rotational quantum number. The degeneracy of states, $g_J$ is given by
 
 \begin{equation}
    g_J = (2J + 1)(2I + 1)
    \label{eq:degen}
\end{equation}
where $I$ is the total nuclear spin, which is 0 for para-H$_2$ (even $J$) and 1 for ortho-H$_2$ (odd $J$). According to the ratio of the partition functions, we expect  OPR$_{\mathrm{eq}}$~$\sim$~3 at temperatures above 200\,K \citep{Burton1992}.

H$_2$ formation on grain surfaces is exothermic \citep{BlackVanDishoeck1987}, with immediate ejection of the newly-formed \hh. The surface-mediated formation reaction is expected to be insensitive to the nuclear spins, leading to a measured OPR=3 for the newly-formed \hh\ entering the gas.

As gas cools in equilibrium, ortho-H$_2$ is converted to para-H$_2$ through H--H$_2$ and H$^+$--H$_2$ reactive collisions, and OPR$_{\mathrm{eq}}$ decreases \citep{Fuente1999,Flower2006}, evolving according to the ratio of the partition functions in Equation~\ref{eq:OPR_eq}. Thus, cold gas below $\sim$\,80\,K is dominated by the para-state \citep[OPR$\,<\,$1][]{Burton1992}.

\subsubsection{Thermal Disequilibrium of the OPR}
\label{sec:nonequilOPR}
The observed OPR provides a window into the recent history of the molecular gas \citep{Neufeld2006}, including \hh\ formation on grain surfaces (with OPR often assumed to be 3:1), \hh\ photodissociation (with preferential photodissociation of para-\hh\ when OPR$>$1) and slow ortho-to-para conversion in the gas (and on grain surfaces). 

At high temperatures ($\gtrsim 300\,$K), spin conversion reactions proceed $\sim$3 times faster for para-to-ortho-\hh\ transitions relative to ortho-to-para-\hh\ transitions \citep{LeBourlot1999, GL2021}.  At lower temperatures this remains the case, with proton exchange reactions dominating species conversion \citep{GL2021,Lique2014}. Below $\sim75\,$K the endothermic para-ortho conversion reaction (e.g. $J=0\rightarrow1$) slows significantly \citep{Lique2014}.

The cosmic ray ionization rate (CRIR) impacts the timescale for spin conversion via collisions with protons since a higher CRIR increases the number of free protons in the gas enhancing conversion \citep{FlowerWatt1984}. For a CRIR $\zeta$=10$^{-17}$\,s$^{-1}$, \citet{FlowerWatt1984} find that the timescale for the OPR and the kinetic and rotational temperatures of the gas to come into thermal equilibrium is $\sim$\,10$^6$\,years for densities n(H$_2$)\,$\sim$\,10$^2$--10$^3$\,cm$^{-3}$ and $\sim$\,10$^7$\,years for densities n(H$_2$)\,$\sim$\,10$^4$--10$^5$\,cm$^{-3}$ where the free proton density is lower. \citet{vanderTak2016} finds $\zeta$=(6-80)$\times$10$^{-17}$\,s$^{-1}$ for M82 and thus the timescale for ortho-\hh\ to para-\hh\ conversion through collisions with protons is expected to be shorter. For example, \citet{Pagani2013} finds that the timescale of ortho-to-para-\hh\ conversion via collisions with protons to be of order 0.1\,Myr for $\zeta$=10$^{-16}$\,s$^{-1}$ and gas density of 2\,$\times$\,10$^4$\,cm$^{-3}$.

Inter-species conversion by collisions with H or \hh\ has a substantial activation energy leading to a strong temperature dependence of the reaction rate coefficient \citep{Lique2012, SR1965}. As a result, para-\hh\ to ortho-\hh\ reactions with H atoms will only raise the observed OPR for the fraction of the gas with kinetic energies exceeding the activation barrier of $E/k$ $\sim$\,3900\,K. A significant increase in the observed OPR therefore requires gas kinetic temperatures of order 1000\,K of higher \citep{LeBourlot1999} on timescales exceeding the temperature-dependent timescale for OPR thermalization. Above 300\,K, where inter-species conversion reactions by collisions with H become significant, the gas must be warm for $\sim$~0.1\,Myr for thermalization of the OPR, as determined from the reaction rate coefficients of \citet{Lique2012}, assuming gas density $n$=10$^4$\,cm$^{-3}$. This timescale decreases as the gas temperature increases.

In summary, at the typical temperature of warm \hh\ gas, the processes which can bring ortho- and para-\hh\ into equilibrium are relatively slow, ranging in timescales from 0.1--10\,Myr, depending on temperature, CRIR, and other physical conditions. Non-equilibrium OPRs thus provide information about the thermal history of the gas in situations where heating or cooling events occur more rapidly than the inter-species conversion timescale. In particular, during rapid heating the OPR can remain substantially frozen, retaining information about the thermal history of the once cooler gas. 

\section{MIRI/MRS Data reduction} \label{sec:data}
JWST MIRI/MRS observations of the M82 nuclear starburst were taken as part of the JWST Cycle 1 GO Program 1701 (PI: A. Bolatto) between December 31 2023 and January 1 2024 (UT). The data were obtained from the Mikulski Archive for Space Telescopes at the DOI: \dataset[10.17909/ta3c-0426]{http://dx.doi.org/10.17909/ta3c-0426}. The observations used a 4-point dither pattern with 30 groups per integration and the FULL sub-array with the FASTR1 readout pattern. Each exposure had a duration of 333.004s. The data consist of seven pointings, with observations covering 4.9--27.9 µm from all four integral-field units (channels 1-4). Observations were reduced using all three stages of the JWST Calibration Pipeline with version 1.19.1 \citep{Bushouse2025} and the reference file \texttt{jwst\_1413.pmap}. Changes that were made from the default pipeline to reduce the impact of artifacts are outlined below. 

Detector and pipeline reduction artifacts were identified in the extracted spectra. On either side of the [\ion{Ar}{2}]~6.99\,µm\ emission line profile pronounced redshifted and/or blueshifted absorption ``valleys'' were present. This absorption is an artifact from the \texttt{straylight} step in the calwebb\_spec2 pipeline, which aims to correct for contamination of extracted spectra by the MIRI cross artifact feature in Channel 1 \citep{Gaspar2021, Argyriou2023}. This correction oversubtracted the continuum around the bright 6.99\,µm [\ion{Ar}{2}] line, causing the false absorption features. To fix this, the default reference file for the \texttt{straylight} step was replaced with \texttt{mrsxartcor\_hack.fits} (D. Law, priv comm.), which performs a milder continuum subtraction. 

To reduce the impacts of fringing on extracted spectra, the \texttt{residual fringing} step of the calwebb\_spec2 pipeline was enabled. 

Pixel-based background subtraction was implemented using dedicated background observations as the method of background subtraction for its capabilities in identifying and masking bad pixels from the detector. Background observations were taken on November 16, 2024 (UT) at RA=09:50:43.000, $\delta$=+69:49:50.00. While the background observations were taken months after the science observations, the time delay is not of much concern due to the brightness of the M82 nuclear starburst.

The default settings of the calwebb\_spec3 pipeline were run in order to build the mosaic of the seven pointings.

\begin{figure}
    \centering
    \includegraphics[width=0.9\linewidth]{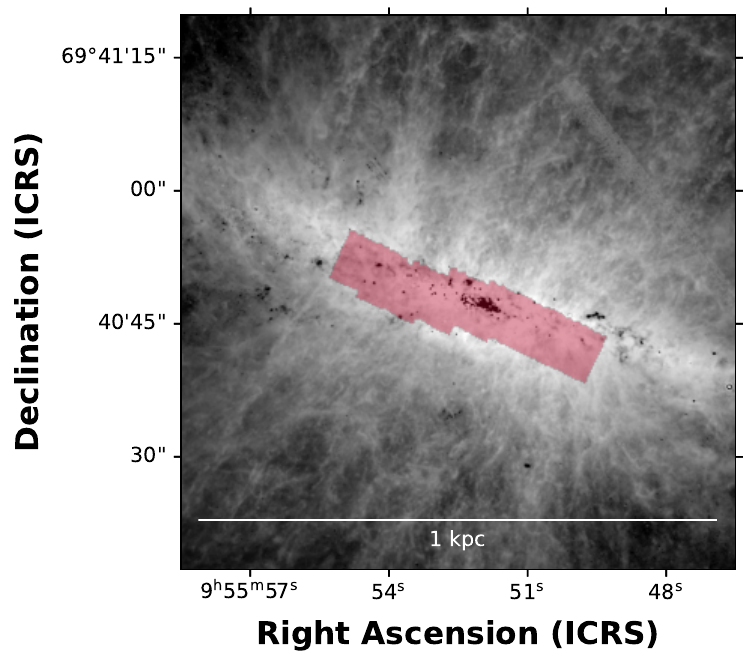}
    \caption{The MIRI/MRS channel 3 mosaic footprint (red) overlaid on the 3.3 µm polycyclic aromatic hydrocarbon map from \citet{Bolatto2024}. The MIRI/MRS data trace the nuclear starburst in M82.}
    \label{fig:FOV}

\end{figure}

\subsection{Spectral Extraction}
\label{subsec:extraction}
The field of view of the mosaic of the MIRI/MRS data is shown in red in Figure~\ref{fig:FOV}, overlaid on  a 3.3\,µm polycyclic aromatic hydrocarbon map from \citet{Bolatto2024}. The MIRI/MRS data trace the center of the nuclear starburst in M82.

Spectral extractions are performed across the total mosaicked area to assess spatial variations in heating conditions. To reduce the impacts of varying spatial resolution, the data are convolved to the resolution of the S(1) transition at 17.04 µm by approximating the point-spread functions (PSF) as Gaussian profiles with radii growing according to the FWHM of the MIRI MRS PSF \citep[Equation 1 of][]{Law2023}. Apertures for spectral extractions are squares with 1 arcsecond side lengths, corresponding to a projected distance of $\sim$\,17.5$\times$17.5~pc$^2$, and are half-overlapping with one another. Square-shaped apertures allow for the construction of maps of spatially varying parameters (see \S\ref{subsec:spatialvariations}). These extractions produced 493 spectra. Spectra are stitched together assuming multiplicative offsets between subbands, holding the flux in Channel 3 long as truth. Offsets were generally $<$~10\%, but $\sim$~11\% of spectra required offsets exceeding 30\% between channels.

\begin{figure*}
    \centering
    \includegraphics[width=0.9\linewidth]{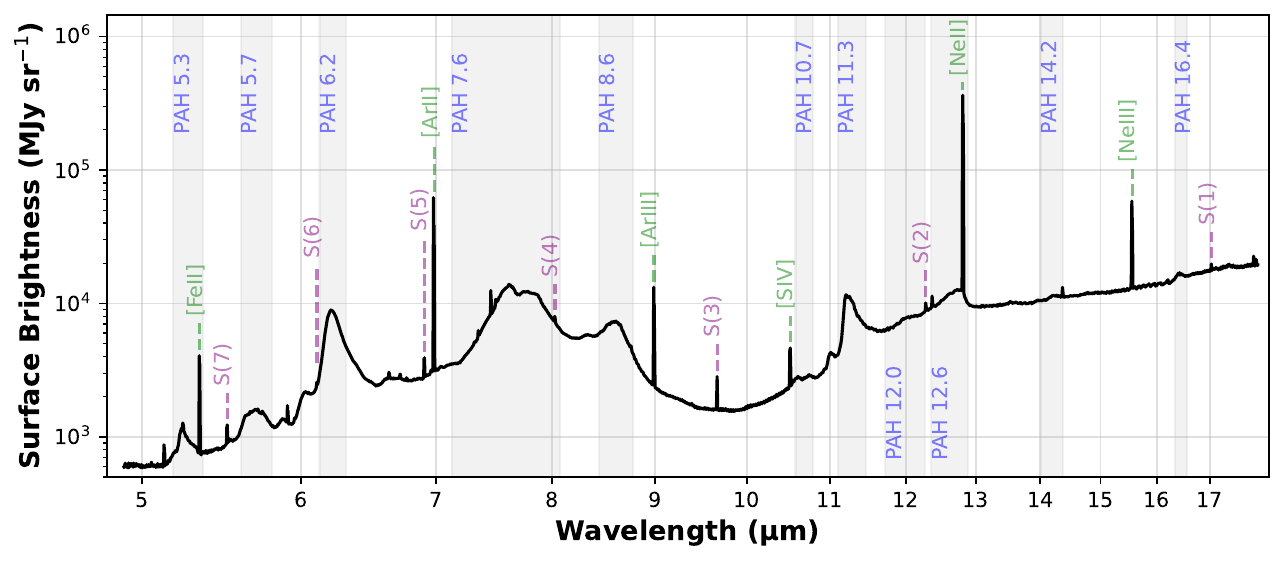}
    \caption{Averaged surface brightness across the total mosaicked area of the M82 nuclear starburst corresponding to the red footprint in Figure~\ref{fig:FOV} with prominent emission lines and dust features labeled.}
    \label{fig:totalSpectrum} 
\end{figure*}

The average spectrum across the full mosaicked area of the M82 nuclear starburst is presented in Figure~\ref{fig:totalSpectrum}. The spectrum shows prominent emission lines and dust features, such as emission from polycyclic aromatic hydrocarbons and absorption from silicates.

\begin{figure*}
    \centering
    \includegraphics[width=0.9\linewidth]{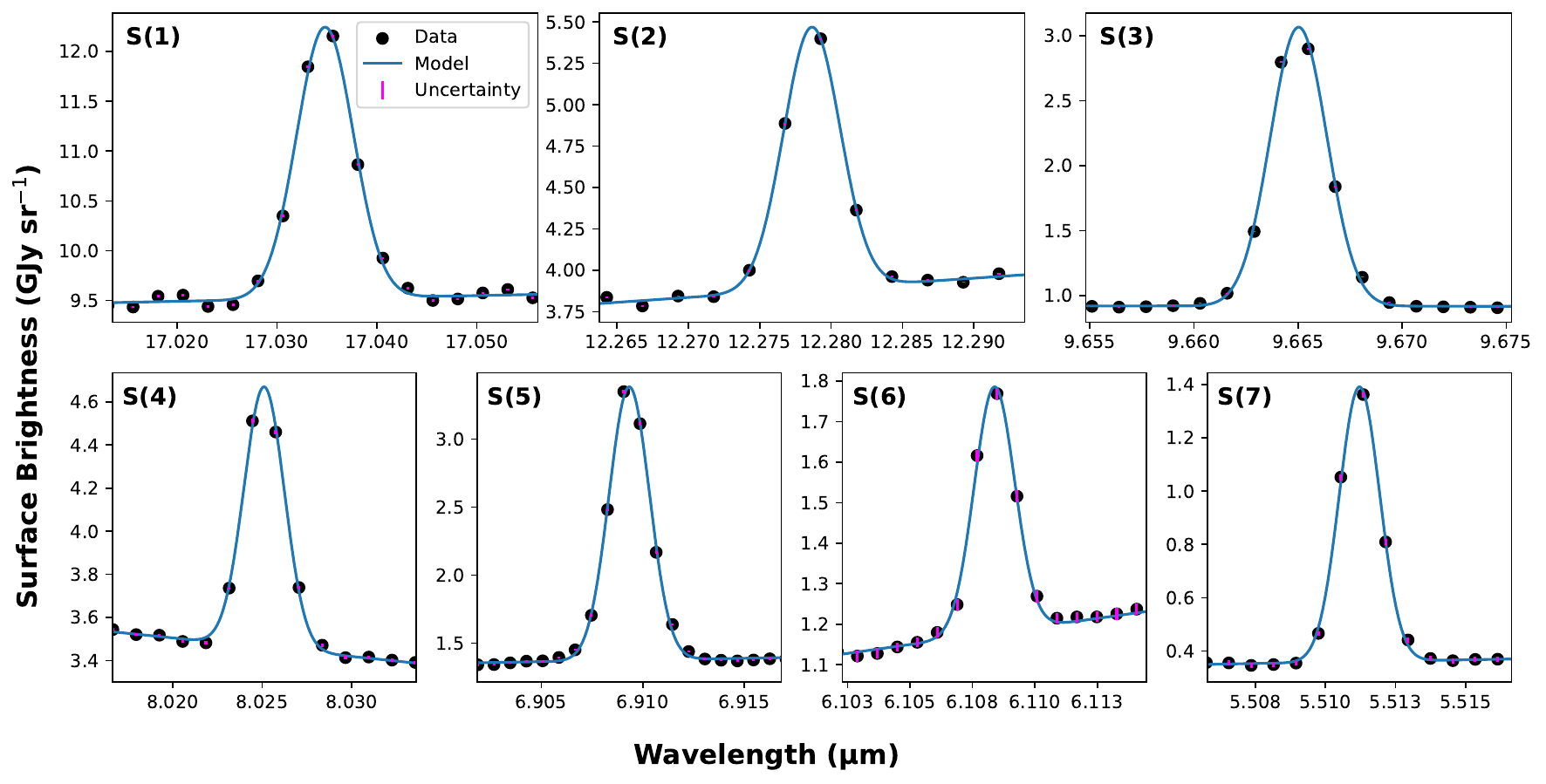}
    \caption{Emission from pure rotational \hh\ transitions S(1)--S(7) (black circles) fit with a model (blue line) composed of a Gaussian profile and a line to model the underlying continuum from a sample extracted spectrum in the M82 nuclear starburst. This region is marked with a magenta star in Figure~\ref{fig:S1SB}. Uncertainties from the pipeline are shown in magenta. See \S\ref{subsec:extraction} for more details on the spectral extraction and model.}
    \label{fig:linefits}
\end{figure*}

Gaussian profiles are fit to the H$_2$ S(1)--S(7) pure rotational transitions along with a line to model the underlying continuum to calculate line fluxes which are used for subsequent analysis. S(8) is not detected. Figure~\ref{fig:linefits} shows \hh\ S(1)--S(7) line fits for one of our extracted spectra, with the location of this spectral extraction marked by the magenta star in Figure~\ref{fig:S1SB}.  

\subsection{Modeling Velocity Dispersion}
\label{sec:velocitydispersion}

\begin{figure}
    \centering
    \includegraphics[width=0.9\linewidth]{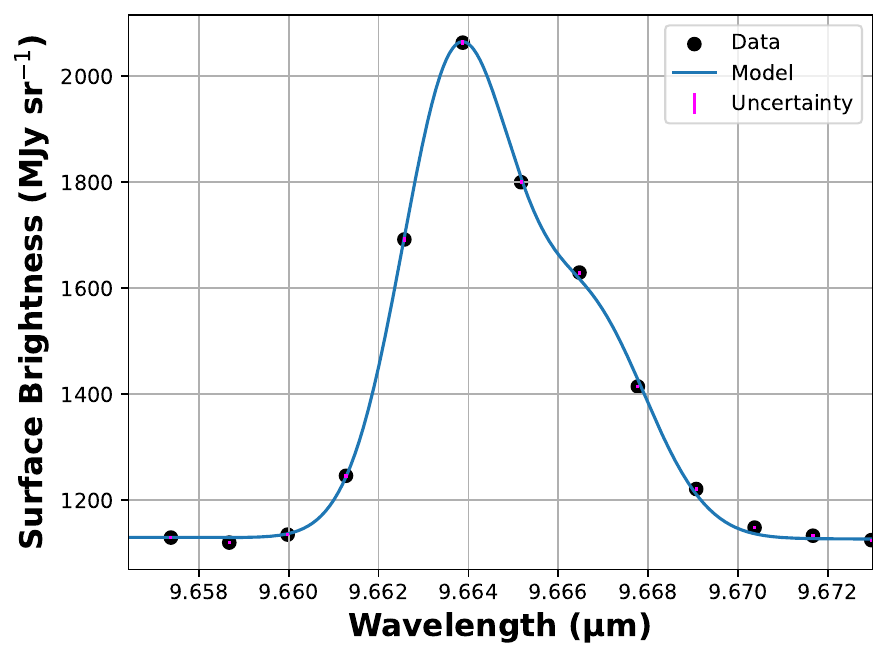}
    \caption{Sample emission from the S(3) pure rotational transition of \hh\ demonstrating velocity broadening (black) with the model shown in blue. Uncertainties from the pipeline are shown in magenta. A Gaussian profile is fit to each velocity component as discussed in \S\ref{sec:velocitydispersion}. The two velocity components are separated by 90.6\,\kms.}
    \label{fig:twocompfit}
\end{figure}

A defining characteristic of M82 is its prominent galactic wind (\citet{Walter2002, Veilleux2009, Bolatto2024}, S. Lopez et al. 2025, submitted), which is powered by the numerous massive star clusters in the central burst \citep{McCrady2007,Mayya2008,Levy2024}.
Such outflows and winds can cause the molecular gas to move at a range of speeds due to turbulence which manifests as broadening in the emission line profiles. Additionally, M82 is an edge-on galaxy, so the telescope beam may cover several components of gas that could be moving at different radial velocities \citep{Krieger2021}. Accounting for line broadening is thus important for accurate measurement of emission line fluxes.

To account for broadening in the line profiles, we allow the fitted width of the Gaussian profile to vary by up to twice the width of an unresolved spectral line (i.e., the instrumental width, see \citealt{Labiano2021}), amounting to $\sim$~70--100\,\kms. By modeling the instrumental profile of the pure rotational transitions of \hh\ with a Gaussian profile and allowing its width to increase to simulate velocity broadening, we determine that velocity dispersions can be reliably measured down to $\sim$~50\,\kms\ for transitions S(3)--S(6). For transitions S(1), S(2), and S(7), the threshold for reliable measurement is between 70-100\,\kms. These values are determined such that the fitted, broadened velocity dispersion is within 5\% of the input broadening. 

Fewer than 9\% of sight-lines demonstrate detectable velocity broadening in at least one pure rotational \hh\ transition and thus 91\% of sight-lines have linewidths consistent with the instrumental linewidth. We only detect reliable velocity broadening in transitions S(3)--S(6) where $\sigma_v~>$~50\,\kms. $\sim$1\% of sight-lines contain H$_2$ emission lines with two velocity components, with peaks separated by $>$\,90\,\kms. To model both components, we use two Gaussian profiles. A sample two-velocity component line fit is shown in Figure~\ref{fig:twocompfit}, where the velocity components are separated by 90.6\,\kms. Both components have widths consistent with the instrumental width. Regions with increased velocity dispersions in the S(3) transition are marked by the red and salmon contours in the top panel of Figure~\ref{fig:allmaps}, which correlate spatially with the velocity dispersions measured in transitions S(4)--S(6).

\section{Disequilibrium Model}
\label{sec:ksimodel}
\citet{TS2016} (hereafter,\defcitealias{TS2016}{TS2016} \citetalias{TS2016}) use a power-law temperature distribution to model the upper level column densities of \hh\ observed by Spitzer in a number of different galaxies, with the assumption that the total column density within temperatures $T$ and $T$\,+\,$dT$ is proportional to $T^{-n}$, where $n$ is the power-law index. The power-law is integrated over a range of temperatures to obtain an estimate of the total column density of H$_2$ in that temperature range. We take the lower and upper temperatures in the integral to be $T_l$\,=\,100\,K and $T_u$\,=\,2000\,K from \citetalias{TS2016}. \citetalias{TS2016} determines $T_l$ as the lower-limit where the power-law could be reliably extrapolated down to to recover an estimate for the total molecular gas mass and $T_u$ as the temperature above which the value of recovered molecular gas mass converges. $T_l$ is not a significant contributor to our model as long as it is low enough (we are not attempting to recover the total molecular gas mass here).

We express the column density of the upper energy level of a given pure rotational emission line $J+2\rightarrow J$, without making any assumption of whether the OPR obtains its equilibrium value, as

\begin{equation}
    N(J) = 
    \begin{cases}
       m \int_{T_l}^{T_u} \dfrac{1}{\rm 1 + OPR} \dfrac{g_J e^{-E_J/kT_\mathrm{{rot}}}}{Z_p(T_\mathrm{{rot}})}T_{\mathrm{rot}}^{-n}dT_\mathrm{{rot}} & \text{$J$=even} \\
        m\int_{T_l}^{T_u} \dfrac{\rm OPR}{\rm 1 + OPR}\dfrac{g_J e^{-E_J/kT_\mathrm{{rot}}}}{Z_o(T_\mathrm{{rot}})}T_{\mathrm{rot}}^{-n}dT_\mathrm{{rot}} & \text{$J$=odd}
    \end{cases}
    \label{eq:levelpop}
\end{equation}

where $T_{\mathrm{rot}}$ is the rotational temperature of the gas\footnote[1]{We note that $T_{\mathrm{rot}}$ and the kinetic temperature of the gas, $T_{\mathrm{kin}}$ are coupled. If, however, densities are below the collisional critical density \citep[$\sim2\times10^2$\,cm$^{-3}$ at $T_\mathrm{kin}\sim 500$\,K,][]{LeBourlot1999}, collisions may not occur fast enough to maintain equilibrium between $T_{\mathrm{rot}}$ and $T_{\mathrm{kin}}$ and non-equilibrium kinetic-rotational effects may be seen, particularly at high $J$.  Our models focus exclusively on rotational temperature $T_\mathrm{rot}$.} and $m$ is the scaling coefficient which is related to the total column density, $N\mathrm{_{tot}}$, through 
\begin{equation}
    m = \frac{(n-1)N_\mathrm{{tot}}}{T_l^{1-n} - T_u^{1-n}} .
\end{equation}

Note that when the OPR is equal to its equilibrium-value ($\mathrm{OPR}_\mathrm{eq}$, Eq.~\ref{eq:OPR_eq}), Eq.~\ref{eq:levelpop} reduces to a single integral: Eq.~3 of \citetalias{TS2016}.

Assuming a power-law temperature distribution of molecular gas as traced by \hh, two other effects may alter the observed excitation ladders: attenuation by dust, primarily impacting the S(3) transition at 9.665\,µm due to the strong resonance of silicate dust absorption centered at 9.7\,µm, and the effects of non-equilibrium heating conditions. We discuss how these two effects are accounted for in our model below.

\subsection{Extinction}
\label{subsec:ext}
M82 has significant dust extinction, with an A$_V$ ranging from $\sim$\,3--24~mag on parsec-scales in the central starburst region \citep{Levy2024}, decreasing the line intensity of S(3) relative to the other lines. Lai et al., in preparation, uses observations of pure rotational \hh\ emission in several galaxies from Spitzer and JWST to show that the attenuation curve can be constrained directly with the deficit in the column density of S(3) relative to the other transitions. 

To account for the differential effects of attenuation on specific \hh\ rotational emission lines, the \citet{Gordon2023} extinction curve is implemented in the power-law model \citep{GCC2009,Fitzpatrick2019,Gordon2021,Decleir2022} . We assume a screen geometry and $R_V=3.1$ for M82 \citep{Hutton2015}. Silicate extinction centered around 9.7\,µm primarily impacts the S(3) transition at 9.665\,µm and thus this extinction curve has been normalized at the wavelength of the S(3) transition. To apply extinction, the column density as given in Equation \ref{eq:levelpop} is multiplied by a factor of $e^{-k(\lambda)\mathrm{\tau_{Si}}}$, where k$(\lambda)$ is the relevant extinction curve normalized at 9.7\,µm (such that k(9.7\,µm)=1), and $\mathrm{\tau_{Si}}$ is a fitted parameter, related to the optical depth of the silicate feature at 9.7\,µm. 

\subsection{Disequilibrium Parameter: $\xi$}
\label{sec:diseq_param}
We introduce the parameter $\xi$ to model the observed disequilibrium heating in M82. $\xi$ is defined as the ratio of the observed OPR, to the equilibrium value, $\mathrm{OPR_{eq}}$, as given by

\begin{equation}
    \xitr = \frac{\mathrm{OPR}}{\mathrm{OPR_{eq}}} = \mathrm{OPR} \times \frac{Z_p(T_\mathrm{{rot}})}{Z_o(T_\mathrm{{rot}})}
    \label{eq:ksi}
\end{equation}

In equilibrium, $\xi$\,=\,1 and OPR$=OPR_\mathrm{eq}$ (see Equation~\ref{eq:OPR_eq}, \S\ref{sec:bkg}). A value of $\xi \ne 1$ indicates disequilibrium between the gas temperature as implied by the observed OPR and the gas temperature based on rotational level populations (see \S\ref{sec:nonequilOPR}). 

\subsubsection{Temperature-dependent $\xi$}
\label{subsec:temp_dep_ksi_model}
As discussed in \S\ref{sec:nonequilOPR}, initially non-equilibrium OPRs can equilibrate rapidly for gas temperatures of order 1000\,K or higher \citep{LeBourlot1999, Lique2012}. Further, UV pumping of the high $J$ levels in addition to collisional excitation may provide extra heating to the gas and higher gas temperatures generally have higher spin conversion reaction rates. 
This implies that warmer molecular gas along the line of sight, which dominates the higher $J$ transitions of H$_2$, may equilibrate more rapidly than the cooler gas emitting the bulk of the lower $J$ line luminosity. This temperature dependence of the equilibration timescale can explain the ``damping'' zig-zag shape observed in some \hh\ SLEDs, as discussed in \citet{Neufeld2006}. 

Previous work by \citet{Neufeld2009} models the OPR as a function of both temperature and time in proto-stellar outflows. They found the OPR of warmer gas to be closer to its equilibrium value compared to cooler gas and quantified the amount of time that the gas has been warm. Given the complexity of heating environment and history in galactic regions spanning $\gtrsim10$\,pc, we take a different approach, introducing a temperature dependence of the disequilibrium parameter $\xi$. The column density as given by Equation~\ref{eq:levelpop} is modified to include the effects of high-temperature equilibration, which for simplicity we take to occur instantaneously above $T_{\mathrm{eq}}\,=$\,1000\,K. Other values of $T_{\mathrm{eq}}$ are tested and discussed in \S\ref{subsec:Teq}. Equation~\ref{eq:ksi} is modified as:

\begin{equation}
    \xitr =
    \begin{cases}
             \frac{\mathrm{OPR}}{\mathrm{OPR_{eq}}} = \mathrm{OPR} \times \frac{Z_p(T_\mathrm{{rot}})}{Z_o(T_\mathrm{{rot}})} & T_l \le T_\mathrm{{rot}} < T_{\rm eq} \\
            1 & T_{\rm eq} \le T_\mathrm{{rot}}.\\
    \end{cases}
    \label{eq:xi_Trot}
\end{equation}

Next, we define a function which generalizes the power-law integral of the column density in Equation~\ref{eq:levelpop}:

\begin{multline}
    S_J(n,\xitr) =   \\
     \int_{T_l}^{T_{u}} \frac{e^{\frac{-E_J}{kT_\mathrm{{rot}}}}T_\mathrm{{rot}}^{-n}dT_\mathrm{{rot}}}{Z_p(T_\mathrm{{rot}}) + \xitr Z_o(T_\mathrm{{rot}})} \times
    \begin{cases}
        1 & \text{$J$=even} \\
        \xitr & \text{$J$=odd}.\\
    \end{cases}
    \label{eq:Nksi_eq_general}
\end{multline}

\noindent Note the additional factor of $\xi$ in Eq.~\ref{eq:Nksi_eq_general} for $J$=odd.  It follows from Equations~\ref{eq:levelpop} and \ref{eq:Nksi_eq_general} that 

\begin{equation}
    \frac{N(J)}{g_J} = m S_J(n,\xitr).
\end{equation}

\noindent The total column density may be obtained by modifying Eq.~3 of \citetalias{TS2016} to 

\begin{equation}
    N_{tot} = \frac{4\pi F_J\lambda(T_l^{(1-n)} - T_u^{(1-n)})}{Ahc\Omega (n-1)g_J S_J(n,\xitr)}
\end{equation}

\noindent Where $F_j$ is equal to Eq.~8 of \citetalias{TS2016}. The observed column densities as function of $J$ are then expressed as:

\begin{equation}
    \frac{N_{obs}(J)}{g_J} = m\,e^{-k(\lambda)\mathrm{\tau_{Si}}}S_J(n,\xitr) 
                 \label{eq:Nobs}
\end{equation}

We use the Python package \texttt{scipy curve\_fit} to minimize Equation \ref{eq:Nobs} to find the best-fit values for fitted parameters $n$, $\xi$, $m$, and $\tau_{\mathrm{Si}}$, which are simultaneously fit. Fitting for four parameters, we require a minimum of five ground-state rotational transitions of H$_2$ to be detected with S/N\,$>$\,3 within an aperture for the fit to proceed, with 413 of the 493 spectra satisfying this criterion. This constraint removed all regions containing \hh\ emission lines with two velocity components, as mentioned in \S\ref{subsec:extraction}. Other regions removed are located on the edges of the MIRI/MRS spatial coverage.

\begin{figure}
    \centering
    \includegraphics[width=0.9\linewidth]{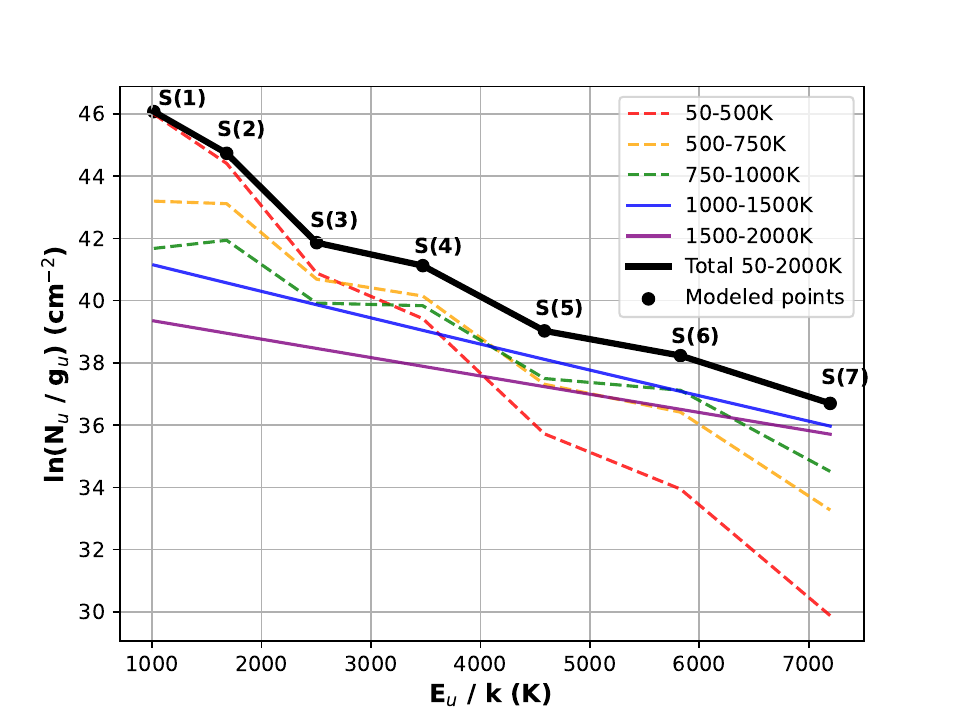}
    \caption{Behavior of the temperature-dependent $\xi$ model (Equation~\ref{eq:Nobs}) across five example gas temperature bins for a modeled spectrum with a power-law index of the temperature distribution, $n$=4.8 and $\xi$=0.35 (1.0) for gas temperatures below (above) 1000\,K. Pure rotational transitions S(1)--S(7) are labeled. As the gas temperature increases, the shape of the resulting SLED (S(1)--S(7)) approaches a flat line. The flat, solid lines for the two highest temperature bins demonstrate OPR thermalization for gas temperatures $>$\,1000\,K. The solid black line shows the resulting disequilibrium model, integrated over the full temperature range of 50--2000\,K.}
    \label{fig:model_sum_cartoon}
\end{figure}

To illustrate the behavior of the temperature-dependent $\xi$ model (\xitr) across different molecular gas temperatures, Figure~\ref{fig:model_sum_cartoon} shows modeled SLED components for five temperature bins with no attenuation but in substantial disequilibrium: $\xi$\,=\,0.35. As the temperature values increase, the slope of the corresponding SLED component becomes shallower and the shape flattens. While steeper lines show the pronounced zig-zag of disequilibrium, 
the flat lines for the two highest temperature bins demonstrate OPR thermalization for gas temperatures $>$\,1000\,K as enforced by the model. The total model is the black, solid curve, showing the resulting damped zig-zag shape.

\begin{figure*}
    \centering
    \includegraphics[width=0.9\linewidth]{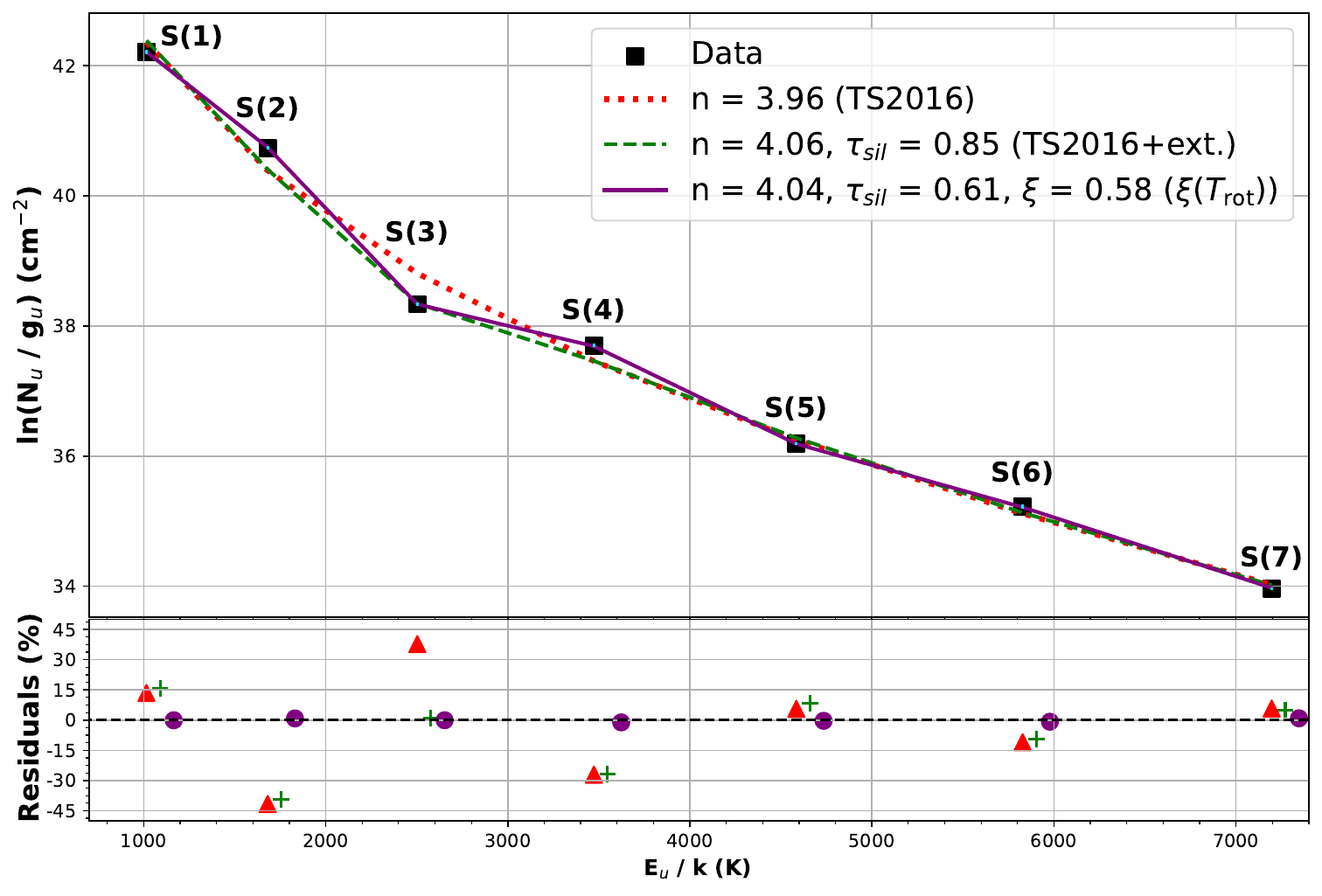}
    \caption{Top panel: Three sample model fits to observed column densities extracted (black squares with uncertainties in cyan) from a 1\arcsec square aperture in the M82 center (region is marked in Figure~\ref{fig:S1SB} by a magenta star). The models are: the \citetalias{TS2016} model (red, dotted curve), \citetalias{TS2016} + extinction (green, dashed curve), and the disequilibrium model (solid, purple curve) (\S\ref{subsec:temp_dep_ksi_model}, Equation~\ref{eq:Nobs}). Fitted parameter values are shown in the legend. Bottom panel: The residuals between the modeled and observed column densities. Points for models with increasing numbers of parameters are offset to the right for clarity. The \citetalias{TS2016} model (red triangles) notably overestimates the column density at S(3), which improves upon the addition of the model parameter $\mathrm{\tau_{Si}}$ (green crosses), expressing the optical depth of silicates. Low energy transitions show a zig-zag shape in their residuals, damping out at higher energy transitions. When the temperature-dependent $\xi$ parameter is introduced, residuals are $<$\,2\% for transitions S(1)--S(7) (purple circles). }
    \label{fig:modelCompsResids}
\end{figure*}

To validate the parameters \ts\, and $\xi$ in the context of disequilibrium heating, the top panel of Figure~\ref{fig:modelCompsResids} models the observed column density from a sample region in M82 using three different models: the \citetalias{TS2016} model is shown by the red, dotted curve, \citetalias{TS2016} + extinction is shown by the green, dashed curve, and the full disequilibrium model with a temperature-dependent $\xi$ (\xitr) is shown in the solid, purple curve. The residuals between the modeled and observed column densities in the bottom panel reveal that the disequilibrium model (purple) is best able to capture both the deficit in S(3) from extinction and the damping zig-zag shape resulting from disequilibrium heating, with residuals $<$2\%. Note that the value of \ts\ is slightly reduced compared to the \citetalias{TS2016} + extinction model, since the S(3) transition strength is reduced both by attenuation and OPR suppression.

\section{Results} \label{sec:results}
Following the method described in \S\ref{subsec:extraction}, we detect bright \hh\ emission from pure rotational transitions S(1)--S(7) across the M82 nuclear starburst. Of the 493 spectral extractions performed across the M82 nuclear starburst (see \S\ref{subsec:extraction}), 413 have detections of at least five pure rotational transitions of \hh\ (with S/N$>$3), used for subsequent analysis of spatial variations in \hh\ excitation. 331 extracted spectra contain detections of the seven pure rotational transitions of interest, S(1)--S(7).

\begin{figure*}
    \centering
    \includegraphics[width=0.9\linewidth]{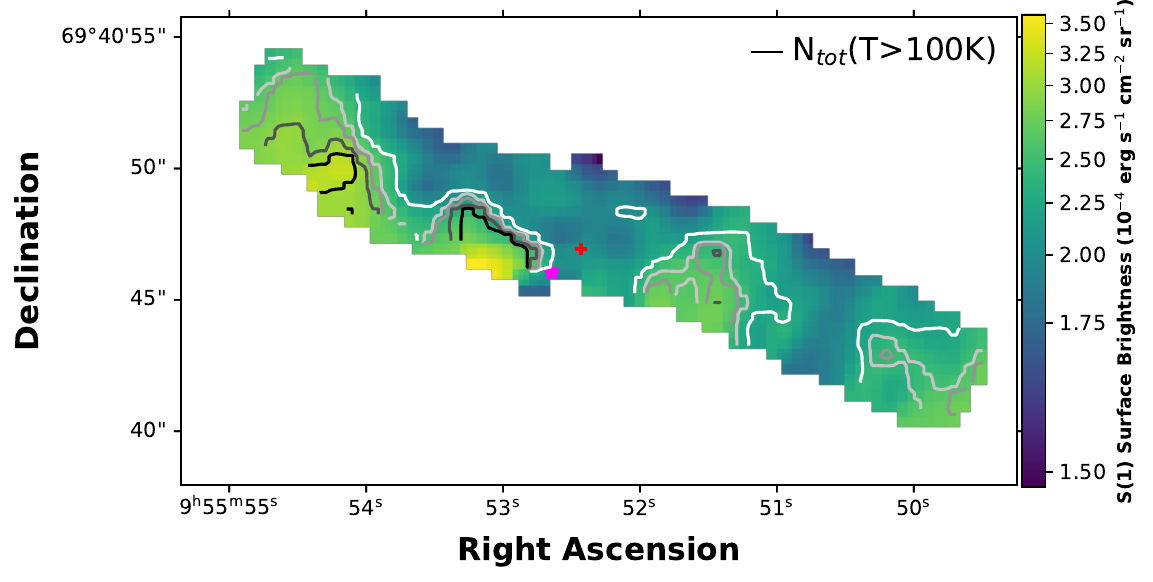}
    \caption{Total surface brightness of the pure rotational transition, S(1), at 17.04\,µm across the M82 nuclear starburst for the 493 extracted spectra. The galaxy center is marked with the red plus sign. Black, gray, and white contours trace the total modeled column density across transitions S(1)--S(7) (6.1$\times$10$^{19}$ (white) -- 7.7$\times$10$^{19}$\,cm$^{-2}$ (black)) for gas temperatures above 100\,K. Regions of increased surface brightness in S(1) generally trace the total column density. The magenta star denotes the location of the spectral extraction for which the \hh\ pure rotational emission line fits are shown in Figure~\ref{fig:linefits}.}
    \label{fig:S1SB}
\end{figure*}

The surface brightness of the pure rotational transition, S(1), at 17.04\,µm is shown in Figure~\ref{fig:S1SB}. We apply our disequilibrium model (\S\ref{subsec:temp_dep_ksi_model}, Equation~\ref{eq:Nobs}) to all extraction regions and transitions with S/N$>$3. From the model fits, we determine the following inner eight deciles of parameter values to be: 4.07--5.09 for the power-law index, $n$, 0.33--0.65 for the disequilibrium parameter, $\xi$, and 0.57--2.0 for the optical depth of silicates, \ts. Black, gray, and white contours trace the total modeled column density across transitions S(1)--S(7). Regions brighter in emission from S(1) generally trace the measured total column density, as expected. 

\subsection{Model Validation through Excitation Diagram Residuals} \label{sec:modelval}
\begin{figure}
    \centering
    \includegraphics[width=0.9\linewidth]{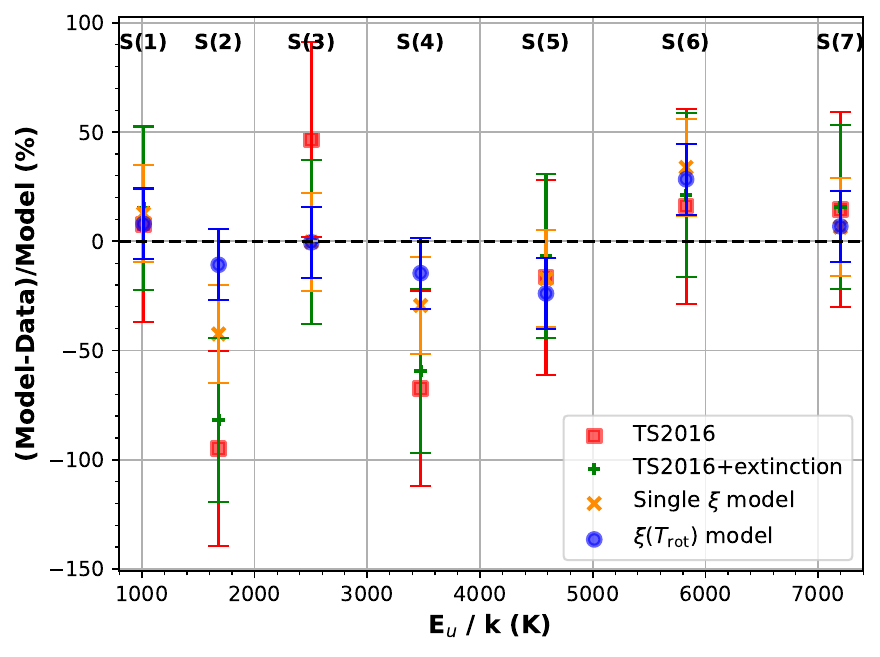}
    \caption{Median residuals between modeled and observed column densities for 331 regions across M82 which contained detections of the S(1)--S(7) pure rotational transitions of H$_2$. The inner three quintiles of the fits are shown by the associated bars for the four models tested. The red squares show the \citetalias{TS2016} simple power-law model, which is expanded to account for extinction by dust, shown with the green crosses. The disequilibrium ($\xi$) model (see \S\ref{sec:diseq_param}) is shown with orange $\times$'s for a single value of $\xi$, and blue circles for the temperature-dependent $\xi$ model. See \S\ref{sec:modelval} for more details on the models. Models that do not account for disequilibrium heating tend to underestimate the column density of S(2) and S(4) most dramatically, reaching values $>$\,100\% of the observed value in several regions. Generally, residuals tend closer to zero as more physically-motivated processes are accounted for in the model.}
    \label{fig:model_residuals}
\end{figure}

To test the necessity of introducing additional parameters to the \citetalias{TS2016} model, we fit the observed column densities to the 331 regions (see \S\ref{subsec:extraction} for spectral extraction details) for which all seven transitions S(1)--S(7) are reliably detected to four different SLED model classes which build upon one another. The resulting median and inner three quintiles residuals of the fits are shown in Figure~\ref{fig:model_residuals}. \citetalias{TS2016} is the simplest model (\S\ref{sec:ksimodel}), employing a power-law to model the temperature distribution of the molecular gas. While capturing the equilibration observed for higher energy transitions, this model (red squares) significantly overestimates the column density at the S(3) transition. To address this discrepancy, the parameter \ts\ is introduced to model attenuation parameterized by the optical depth of silicates (\citetalias{TS2016}+extinction; see \S\ref{subsec:ext}), and the residuals at S(3) approach zero (green crosses). However, both of these models underestimate the column densities at S(2) and S(4). At S(2), \citetalias{TS2016} and \citetalias{TS2016}+extinction underestimate the column density by median values of $\sim$ 95\% and 82\%, which is significant given the median uncertainty on the column density of transitions S(2) and S(4) is $\sim$ 5\%.
At S(4), the column density is underestimated by a median value of $\sim$ 67\% and 59\% for \citetalias{TS2016} and \citetalias{TS2016}+extinction. This leads to a ``damping zig-zag'' shape in the residuals which flattens out moving towards higher energy transitions --- a sign of ortho/para disequilibrium in the lower energy transitions.

To account for the effects of disequilibrium heating, the newly introduced parameter $\xi$ (\S\ref{sec:diseq_param}) is added to the model as a constant value over the full range of temperatures from $\mathrm{T_l}$\,=\,100\,K to $\mathrm{T_u}$\,=\,2000\,K (single $\xi$ model). The orange crosses in Figure~\ref{fig:model_residuals} shows that a constant $\xi$ model reduces median residuals at transitions S(2) and S(4) to $\sim$ 42\% and 29\%, respectively, but the damping zig-zag shape remains in the residuals.

To address the damping, we model excitation with the temperature-dependent $\xi$ model (\xitr; see \S\ref{subsec:temp_dep_ksi_model}, Equation~\ref{eq:Nobs}). In this model, \xitr\, is fixed at its equilibrium value of 1 for gas temperatures $>$\,1000\,K and takes a single fitted value for temperatures $\leq$\,1000\,K. Seen with the blue circles in Figure~\ref{fig:model_residuals}, the median residuals at S(2) and S(4) are much closer to zero ($\lesssim$~15\%). 

Reduced chi-square values are calculated for the four models to further validate the additional model parameters. We find that the temperature-dependent $\xi$ model (\xitr; \S\ref{subsec:temp_dep_ksi_model}) is the best-fit model, with the lowest median and mean reduced chi-square of the four models tested. In order of ascending number of model parameters, the median reduced chi-square values are: 346 \citepalias{TS2016}, 153 (\citetalias{TS2016}+extinction), 65 (single $\xi$ model), and 26 (temperature-dependent $\xi$ model) with 2,3,4, and 4 degrees of freedom, respectively.

The four models presented in Figure~\ref{fig:model_residuals} overestimate the column density at S(6) by a median value of at least $\gtrsim$15\%. Near the S(6) transition, there exists an absorption feature by H$_2$O ice at 6.02\,µm. The prominent 6.2\,µm PAH feature in the M82 spectra makes it difficult to determine if the absorption feature is present. If there is H$_2$O ice absorption, this would cause the column density of S(6) to appear lower than it actually is. This point is explored further in \S\ref{S6_overestimate}, along with other possible explanations for the discrepancy between the modeled and observed column densities at S(6).

\subsection{\ts Validation}
\begin{figure}
    \centering
    \includegraphics[width=0.9\linewidth]{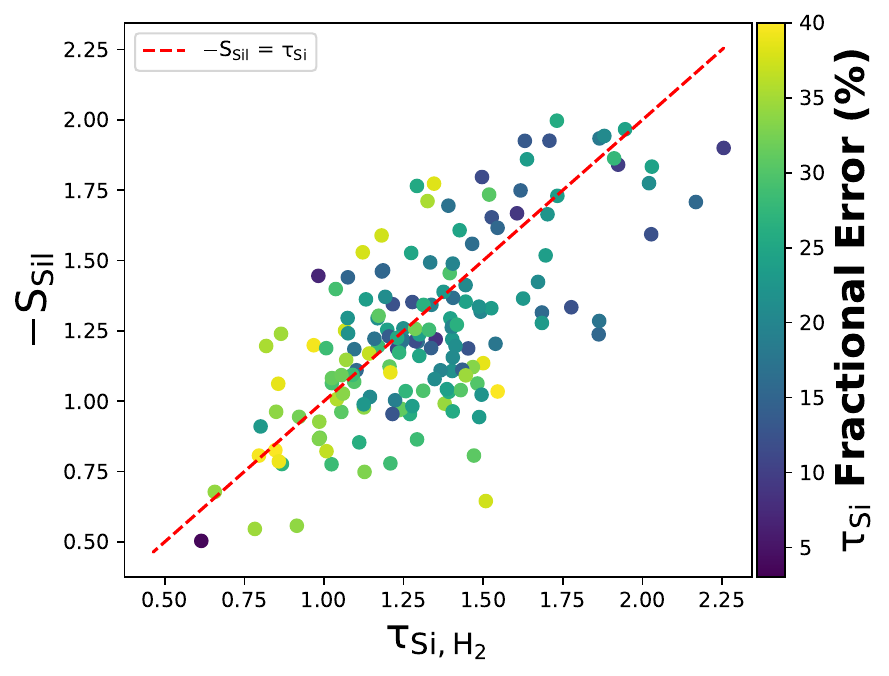}
    \caption{The points show fitted values of \ts\ obtained from our model (\S\ref{sec:ksimodel}) plotted against S$\mathrm{_{Sil}}$ \citep{Spoon2007}. Points are colored by the fractional error of the fitted value of \ts. The dashed, red line indicates the location of -S$\mathrm{_{Sil}}$ = \ts\ (for a screen model).}
    \label{fig:ts_vs_ss}
\end{figure}

To validate the fitted values of \ts\ predicted by our model (see \S\ref{subsec:ext}), we calculate S$\mathrm{_{Sil}}$ \citep{Spoon2007} which estimates the strength of the 9.7\,µm silicate absorption feature and is interpreted as the negative of the silicate optical depth. S$\mathrm{_{Sil}}$ is measured from the continuum based on the depth of the silicate absorption ``trough'' centered around 9.7\,µm, with continuum anchor points at 5.5\,µm and 14.5\,µm \footnote[2]{The silicate strength is calculated according to S$_\mathrm{Sil} = ln\frac{f_\mathrm{obs}(9.7\, \text{\textmu m})}{f_\mathrm{cont}(9.7\, \text{\textmu m})}$, which compares the observed flux density at 9.7\,µm, $f_\mathrm{obs}(9.7\, \text{\textmu m})$, to the estimated flux density of the continuum at 9.7\,µm, $f_\mathrm{cont}(9.7\,\text{\textmu m})$.}. Both methods for silicate optical depth estimation assume a screen geometry for the dust. In Figure~\ref{fig:ts_vs_ss}, it can be seen that \ts\ and -S$\mathrm{_{Sil}}$ are directly related, with scatter around the line of -S$\mathrm{_{Sil}}$ = \ts. We find 95\% confidence intervals of 1.26--1.35 for \ts, and 1.20--1.30 for -S$\mathrm{_{Sil}}$, showing general agreement. We do not expect these methods to produce exactly the same values for the optical depth of silicates because S$\mathrm{_{Si}}$ is measured from the continuum whereas \ts\ relies heavily on the assumed \citet{Gordon2023} extinction curve. 

Our method produces a median fitted \ts\ equal to 1.3~$\pm$~0.3 from the 175 spectra where errors on the fitted value of \ts\ are less than 40\%. Similarly, \citet{Beirao2008} finds a median \ts\,=1.34 and Cronin et al. 2026, in prep. determines \ts~=~1.5 in the central region of M82, both with data from \textit{Spitzer} IRS, through the use of PAHFIT spectral decomposition \citep{Smith2007}. The general agreement supports the process of determining \ts\ from SLEDs.

\subsection{Temperature Distribution}
\begin{figure}
    \centering
    \includegraphics[width=0.9\linewidth]{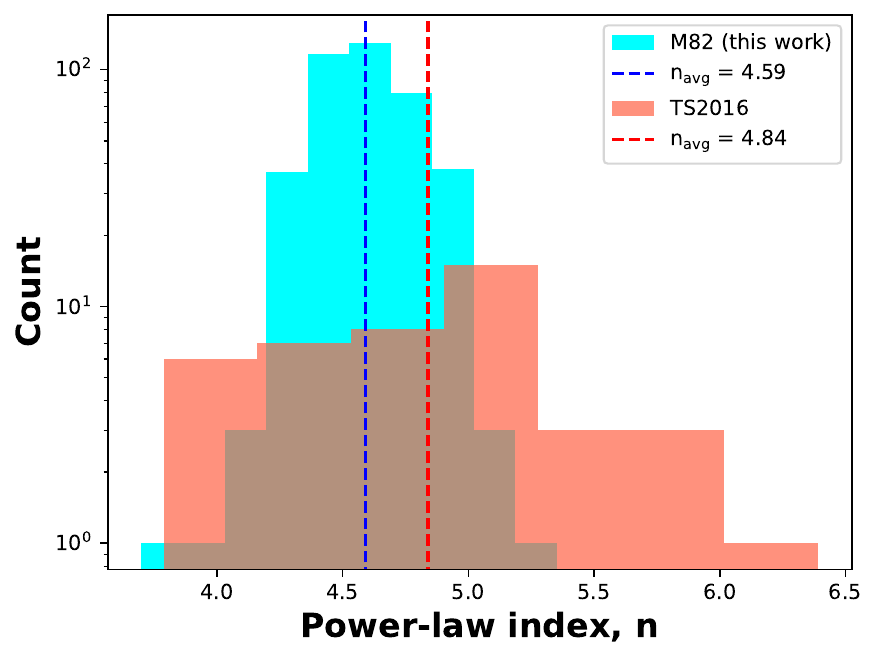}
    \caption{Comparison of the distribution of the power-law index $n$ of the temperature distribution of molecular gas for this work (cyan) and galaxies in \citetalias{TS2016} (tomato). Average values are shown in dark blue and red dashed lines for this work and \citetalias{TS2016}, respectively. Based on the distributions, $n$ measured in M82 is lower, indicating slightly warmer gas temperatures than those in the galaxies in \citetalias{TS2016}.}
    \label{fig:n_hist}
\end{figure}

The power-law index of the temperature distribution provides information about the fractional amount of warm gas present which can help quantify the impact of heating mechanisms such as shocks or UV pumping. Lower power-law indices imply greater contributions of warmer gas, which are expected in shock-heated regions. \citet{NY2008} find that for supernova remnant IC 443, modeling \hh\ column densities with a continuous temperature distribution produces power-law indices ranging from 3--6, and specifically predict $n$\,$\sim$3.8 for C-type bow shocks.

\citetalias{TS2016} determines the power-law index of the temperature distribution of molecular gas for a sample of galaxies with varying environments (dwarf galaxies, radio galaxies, ULIRGS, etc.), which ranges from 3.79 to 6.39. \citet{Mills2017} finds power-law indices, $n$ = 3.22, 2.83 for two regions within the Galactic center circumnuclear disk, indicating large fractions of hot gas.  We find that the temperature distribution of molecular gas as traced by pure rotational \hh\ emission in the nuclear starburst of M82 falls within the range of the \citetalias{TS2016} sample. Figure~\ref{fig:n_hist} compares the values of the power-law index of the temperature distribution in this work to those from \citetalias{TS2016} with the average values presented by the vertical dashed lines. Our work presents a narrower distribution which is expected since we are looking within a $\sim$500~pc-sized region of one galaxy, as opposed to the broader \citetalias{TS2016} sample. Our average power-law index is lower than the \citetalias{TS2016} sample by $\sim$5\% which tells us that on average, the nuclear starburst of M82 has a higher fraction of warm gas, which is not a surprise given the recent starburst.

\subsection{Spatially Varying Parameters} \label{subsec:spatialvariations}
Derived parameters from the modeled H$_2$ line fluxes and column densities are used in the creation of maps to assess spatial correlations. Each spectrum produces a set of fitted parameters which are used as data values in the maps. Because the spectral extraction apertures are half-overlapping, parameter values are averaged in the regions of overlap.

\begin{figure*}
    \centering
    \includegraphics[width=0.9\linewidth]{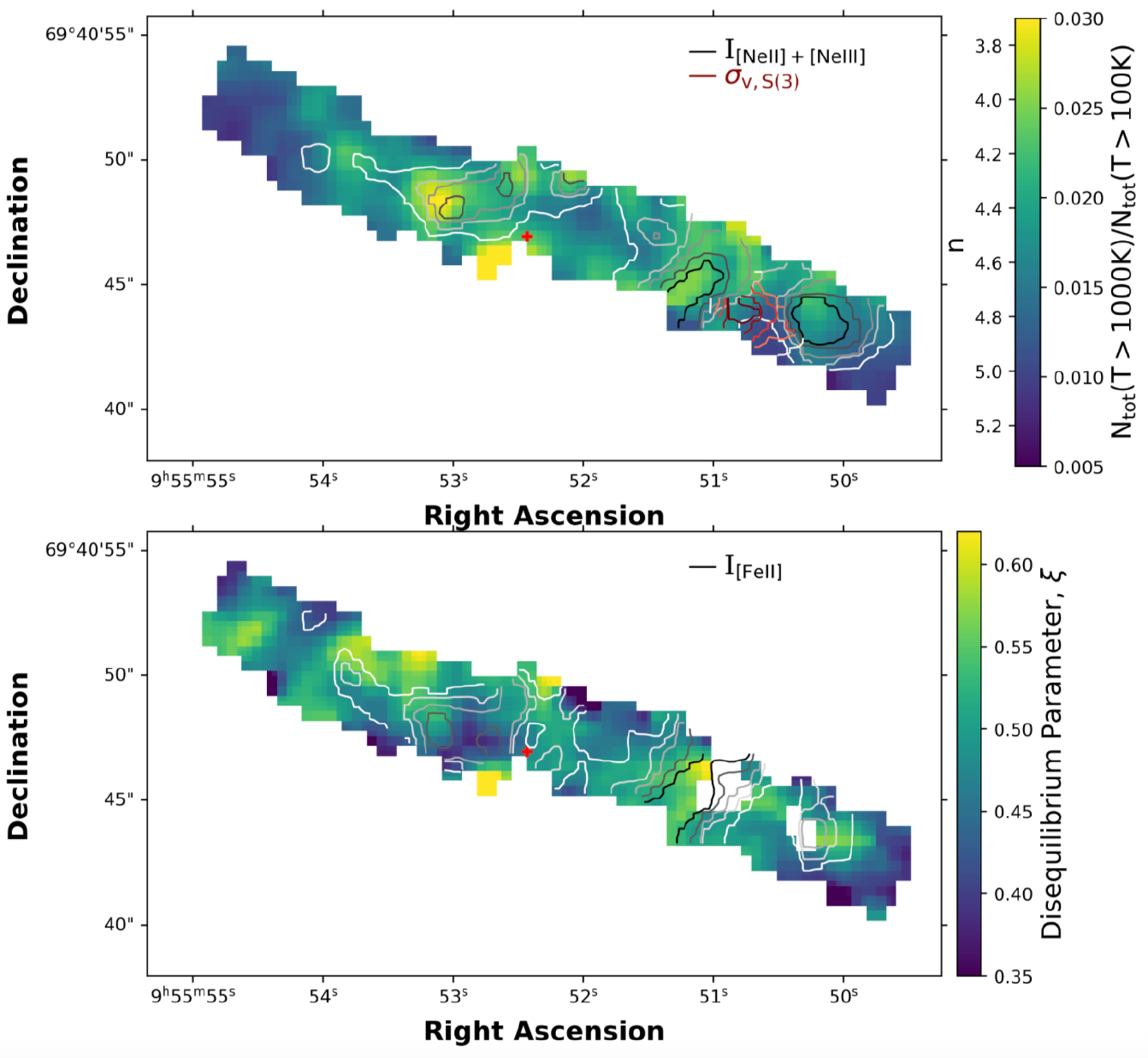}
    \caption{Maps of derived parameters across the M82 center with North facing up. The galaxy center is marked with the red plus sign. Top: Fraction of the column density of hot ($T>$\,1000\,K) to warm ($T>$\,100\,K) molecular gas showing the excitation as a proxy for the power-law index of the temperature distribution, $n$. Darker regions contain more cool gas (steeper power-law index) and lighter regions contain more warm gas (shallower power law index). Black, gray, and white contours show the surface brightness of the star formation tracer, 12.81\,µm\,[\ion{Ne}{2}] + 15.56\,µm\,[\ion{Ne}{3}] \citep{HoKeto2007}, which range from 0.025 (white) -- 0.05 (black) erg\,s$^{-1}$\,cm$^{-2}$\,sr$^{-1}$. The velocity dispersion of the S(3) transition, $\sigma_\mathrm{v,S(3)}$, is shown by the red contours with dispersions ranging from 55.7\,\kms (salmon) -- 65.0\,\kms(dark red).  Bottom: Spatial variations in the disequilibrium parameter, $\xi$. We require model fit parameters to have a fractional error $<$40\%. Darker regions ($\xi$\,$<<$\,1) are retaining effects of disequilibrium heating, which are less dramatic in the lighter regions with increased $\xi$. Black, gray, and white contours trace the surface brightness of the shock tracer, 5.34\,µm [\ion{Fe}{2}] emission, ranging from 0.0008 (white) -- 0.001 (black)\,erg\,s$^{-1}$\,cm$^{-2}$\,sr$^{-1}$.}
    \label{fig:allmaps}
\end{figure*}

\subsubsection{Spatial Temperature Distribution}
Spatial variations in the rotational temperature of molecular gas provide insight into varying heating conditions across the M82 nuclear starburst. The top panel of Figure~\ref{fig:allmaps} shows the hot fraction of gas as defined by $\frac{N_\mathrm{{tot}}(T>1000K)}{N_\mathrm{{tot}}(T>100K)}$ (which maps one-to-one with power law index $n$). Regions with a greater hot fraction correspond to a lower value of the power-law index of the temperature distribution, whereas regions with less hot gas have higher power-law indices, as expected. 
Black, gray, and white contours show the surface brightness of the star formation tracer, 12.81\,µm\,[\ion{Ne}{2}] + 15.56\,µm\,[\ion{Ne}{3}] \citep{HoKeto2007}. Surface brightnesses were determined by fitting Gaussian profiles to the emission lines following the procedure for fitting \hh\ pure rotational transitions as described in \S\ref{subsec:extraction}, and we require a S/N $>$~3 for a line to be detected. The [\ion{Ne}{2}] + [\ion{Ne}{3}] contours correlate spatially (Spearman rank coefficient $\sim$ -0.62, $p$-value $\sim$1.2$\times$10$^{-34}$) with regions containing more warm gas (lower power-law index, $n$) which may indicate that the gas is being radiatively heated through UV pumping by the stars that have formed. The coolest gas is seen around the southwest and southeast edges of the mosaic which generally lack strong [\ion{Ne}{2}] + [\ion{Ne}{3}] emission.

Between the two lobes of [\ion{Ne}{2}] + [\ion{Ne}{3}] emission in the western part of the mosaic, we find an increased velocity dispersion in 45, 24, 15, and 17 sight-lines for transitions S(3)--S(6), respectively, as discussed in \S\ref{sec:velocitydispersion}. The velocity dispersion in the S(3) transition is denoted by the salmon, red, and dark red contours in the top panel of Figure~\ref{fig:allmaps}, with velocity dispersions ranging from 55.7\,\kms (salmon) -- 65.0\,\kms(dark red). The intrinsic velocity dispersion of transitions S(3)--S(6) is broadened by $\sim$ 60~\kms\ on average in these regions, which all correlate spatially with the contours for $\sigma_{\mathrm{v,S(3)}}$. The location of this region between the two lobes of ionized gas may indicate that the lobes are moving in different directions, possibly towards and away from the line of sight, and thus we are observing increased velocity dispersions near their boundaries as a result of either turbulence or two blended velocity components with intrinsically narrow line-widths.

We determine the total warm column density, $N_\mathrm{{tot}}(T>100$\,K)\,=\,(2.0\,$\pm$\,0.1)\,$\times$\,10$^{22}$\,cm$^{-2}$, by integrating Equation~\ref{eq:Nobs} from $T_l$\,=\,100\,K to $T_u$\,=\,2000\,K. From our estimate of $N_\mathrm{{tot}}(T>100$\,K), we find a warm ($T>$100\,K) molecular gas mass of (2.7\,$\pm$\,0.2)\,$\times$\,10$^{7}$\,M$_\odot$, which is $\sim$10.4\% of the total molecular gas mass determined from observations of CO(1-0) in the disk of M82 as determined by \citet{Krieger2021} with the assumption $\alpha_\mathrm{CO}$\,=\,1.0\,M$_\odot$(K\,\kms\,pc$^2$)$^{-1}$.

\subsubsection{Disequilibrium Heating}
\label{sec:diseqheating}
Spatial variations are also found across the M82 nuclear starburst for the temperature-dependent disequilibrium parameter, $\xi$ (Equation~\ref{eq:ksi}), to quantify the degree to which the OPR differs from OPR$_{\mathrm{eq}}$. The bottom panel of Figure~\ref{fig:allmaps} presents a map of $\xi$ for the 363 spectra where errors on the fitted value of $\xi$ are less than 40\%, with contours showing surface brightness of the shock tracer, [\ion{Fe}{2}], at 5.34\,µm. The surface brightness of [\ion{Fe}{2}] was determined following the fitting procedure for \hh\ pure rotational transitions (see \S\ref{subsec:extraction}), requiring a S/N $>$~3 for a line to be detected. In general, $\xi$ does not correlate with  [\ion{Fe}{2}] emission. $\xi$ varies from $\sim$0.33--0.65 with a median value of 0.49 and a standard deviation equal to $\sim$0.06. Values of $\xi$\,$<$\,1 are indicative of disequilibrium heating (\S\ref{sec:diseq_param}) and because we only find $\xi$\,$<$\,1, the observed OPR is \textit{not} in equilibrium with the molecular gas temperature in the M82 nuclear starburst. The relatively low scatter in the fitted value of $\xi$ implies that on average, the observed OPR is roughly half of its equilibrium value at a given temperature. Possible explanations for the widespread suppression of the OPR are explored in \S\ref{sec:model_correlations}.

\section{Discussion} \label{sec:disc}
Armed with the disequilibrium model, we estimate the temperature distribution of warm molecular gas as traced by the pure rotational transitions of \hh, the impact of extinction, and degree of disequilibrium heating in the M82 nuclear starburst. The disequilibrium parameter $\xi$ (\S~\ref{sec:diseq_param}, Eq.~\ref{eq:xi_Trot}) quantifies the differences between the OPR and its equilibrium value given the rotational temperature of the gas, which may be out of equilibrium with each other in situations where molecular gas that has cooled is suddenly heated over timescales shorter than the (temperature-dependent) inter-species conversion timescale ($\xi<1$). 

\subsection{The Thermal History of Molecular Gas} \label{sec:model_correlations}
\begin{figure}
    \centering
    \includegraphics[width=0.9\linewidth]{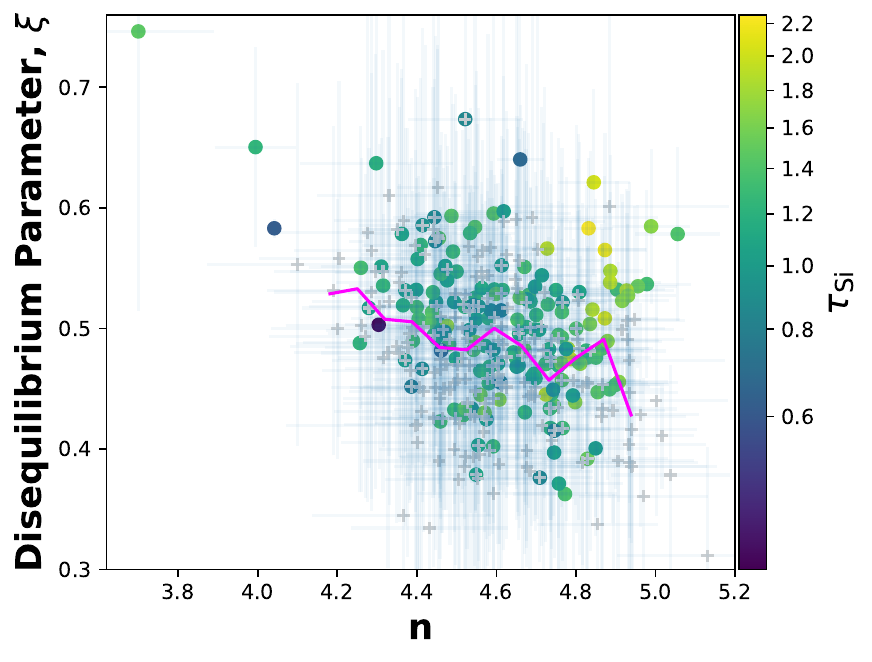}
    \caption{Relations between the best-fit parameters, $n$, the power-law index of the temperature distribution, and $\xi$, the disequilibrium parameter, for fits with fractional errors $<$33\%. The binned medians are shown in magenta. Points are colored by the fitted value of $\mathrm{\tau_{Si}}$. Gray points with cross markers indicate that the fitted value of \ts\ is not well-measured, with an error exceeding 33\% of the fitted value, but $<$100\%. There is a weak anti-correlation between power-law index $n$ and disequilibrium parameter $\xi$, implying that the measured OPR in warmer regions may be closer to its equilibrium-value than in cooler regions.}
    \label{fig:n_vs_ksi}
\end{figure}

To investigate the thermal history of molecular gas in the M82 nuclear starburst with the temperature-dependent $\xi$ model, Figure~\ref{fig:n_vs_ksi} presents the relationship between the disequilibrium parameter, $\xi$, and the power-law index of the temperature distribution, $n$. Points are colored by the fitted value of $\mathrm{\tau_{Si}}$. Gray points with cross markers indicate where fitted values of \ts\ are not well-measured, with an error exceeding 33\% of the fitted value.

A weak anti-correlation between the excitation of the gas as traced by the power-law index of the temperature distribution, $n$, and the disequilibrium parameter $\xi$ emerges with a Spearman rank correlation coefficient of $-$0.26 and $p$-value of 4.1~$\times$~10$^{-6}$. Because the inter-species conversion timescale due to reactive collisions with atomic H rapidly decreases with increasing temperature, it is reasonable to expect that regions with more warm gas (lower $n$) may be equilibrating more quickly than regions with a larger fraction of cooler gas (higher $n$). However, the weakness of the anti-correlation hints at the fact that both the current temperature of the gas and its thermal history together drive the OPR equilibration. A weak positive correlation is observed between $n$ and \ts\, with a Spearman rank correlation coefficient of 0.39 and a $p$-value of 2.7~$\times$~10$^{-6}$, possibly indicating that more dust-embedded clouds have a larger fraction of cooler \hh\ gas, but the complex line of sight geometry complicates this interpretation.

Within the central kpc of M82, there have been two bursts of star formation roughly 8--15~Myr and 4--6~Myr ago with peak star formation rates of 160~M$_\odot$\,yr$^{-1}$ and 40~M$_\odot$\,yr$^{-1}$ \citep{FS2003}. The \textit{lower} measured OPR relative to the equilibrium value for the estimated rotational temperature of the gas in M82 implies that the ortho-to-para abundance ratio is reflecting a cooler prior gas temperature. As the most massive stars that formed end their lives, thermal and kinetic energy is injected into the gas. Thus, if molecular gas cooled after the starburst to an OPR$<$3, reheating from ongoing feedback processes may raise the rotational temperature of the gas on timescales faster than the time required to raise the OPR. Variations in the degree of disequilibrium heating experienced by different parcels of gas may depend on their history of density and temperature.
The analysis of disequilibrium heating in the M82 center as measured by $\xi$ provides a time-average tracer of the thermal history of the varying heating rates, with $\xi$ potentially suppressed after the sudden onset of star formation or other energetic processes.  

\subsection{Turbulent Mixing}
Another physical effect with the potential to globally suppress the OPR in a starburst core is turbulence.  If the timescale for turbulent mixing becomes shorter than the inter-species conversion timescale at relevant densities and temperatures, gas dredged up from cooler regions within clouds could retain lower OPR values, ``remembering'' its cooler past environment. 

The spin conversion timescale is a decreasing function of the temperature of the gas and concentration of H atoms and protons (see \S\ref{sec:nonequilOPR}). For gas kinetic temperatures $\gtrsim\,$300\,K, the rate coefficients of \citet{Lique2012} indicate that spin conversion reactions through collisions with H atoms proceed over timescales $\lesssim$0.1\,Myr, decreasing to the order of hundreds of years for gas as hot as $\sim$500\,K. Below 300\,K where the bulk of the warm \hh\ mass is found, spin conversion timescales for typical cosmic ray heating environments are on the order of a few Myr. As discussed in \S\ref{sec:nonequilOPR}, these timescales may be reduced by up to an order of magnitude in M82 as a result of its increased CRIR.

For a molecular cloud of size $\sim$a few pc with internal velocity dispersions of a few \kms, crossing times of $\sim$1\,Myr, imply that the turbulent mixing timescale should be on the order of a few Myr \citep{HF2012}. In contrast, in molecular gas dominated by turbulent shocks, \citet{Federrath2008} find that the timescale for \hh\ mixing is  $\lesssim\,$0.3\,Myr.

In the shocked, turbulent environment of the M82 nuclear starburst, it is therefore plausible that the spin conversion timescale in warm gas may exceed the turbulent mixing timescale. Within a molecular cloud near a source of UV radiation, turbulent mixing may bring cool \hh\ with a low OPR ($<$3) towards the warmer surface of the cloud, where it heats up rotationally on timescales $\lesssim$\,100\,yr, based on transition probabilities from \citet{BlackDalgarno1976}. Because the timescale for heating an \hh\ molecule is shorter than the spin conversion reaction timescale in this scenario, the molecule will heat up rotationally before the OPR can thermalize with the warmer state of the gas. In less turbulent environments, longer turbulent mixing timescales would imply that cool \hh\ is brought to the warm, irradiated surface more slowly, giving more time for spin conversion reactions to drive the OPR closer to equilibrium in the warmer cloud envelopes.

In summary, the OPR suppression in M82 nuclear starburst may result from the fact that the spin conversion timescale exceeds both the timescale for heating \hh\ rotationally and the fast turbulent mixing timescales, for gas with kinetic temperature $\lesssim$300\,K. Thus, in the highly turbulent environment of the M82 nuclear starburst, the damping zig-zag shape observed in the SLEDs (\S\ref{sec:modelval}) indicate that only the warmest gas remains warm long enough to thermalize. 

Theoretical work exploring the turbulent mixing and OPR thermalization timescales within realistic clouds would be very valuable for validating this scenario. Observationally, the connection between increased turbulence and OPR suppression could be further supported through measurement of increased CO linewidths and therefore lower values of $\alpha_{CO}$ in regions with strong disequilibrium signatures. 

\subsection{Correlations with Ionized and Neutral Gas}

\begin{figure*}
    \centering
    \includegraphics[width=0.9\linewidth]{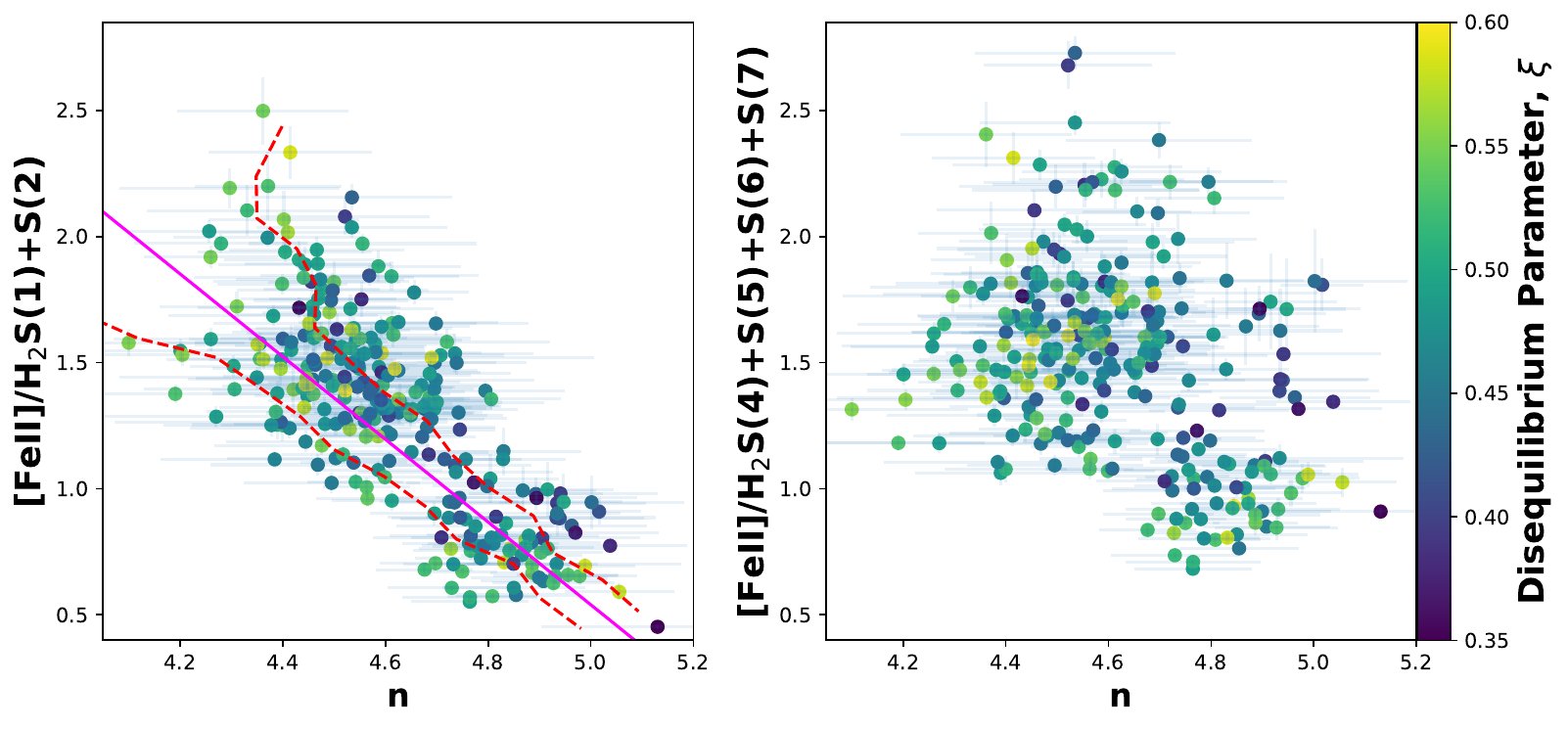}
    \caption{The surface brightness of the [\ion{Fe}{2}] emission line at 5.34 µm relative to the modeled surface brightness of cooler and warmer molecular gas as traced by the pure rotational transitions S(1)--S(2) (left) and S(4)--S(7) (right), respectively, with the power-law index of the temperature distribution, $n$. Points are colored by the fitted value of $\xi$. On the left, the best fit line is shown in magenta with the scatter shown by the dashed red lines, computed as the median perpendicular distance to the best fit line. The anti-correlations that emerge reveal that [\ion{Fe}{2}] emission is enhanced at low $n$ relative to cooler molecular gas.}
    \label{fig:FeIIvsH2vsn}
\end{figure*}

With access to ionic emission lines within the MIRI/MRS coverage, comparisons between the ionized and molecular phases of the gas can provide a more complete picture of the heating conditions within the region. The following emission lines are modeled with Gaussian profiles to measure their surface brightness: [\ion{Ar}{2}], [\ion{S}{4}], [\ion{Ne}{2}], [\ion{Ne}{3}], and [\ion{Fe}{2}]. The surface brightnesses are compared with cool and warm molecular gas as traced by the summations of \hh\ S(1)--S(2) and S(4)--S(7), respectively. 

Relative to the cooler, lower $J$ rotational transitions of \hh\ (S(1) and S(2)), several tracers of ionized gas ([\ion{Ar}{2}], [\ion{S}{4}], [\ion{Ne}{2}], [\ion{Ne}{3}]) show mild to moderate anti-correlations with the power-law index of the temperature distribution, $n$. The [\ion{Fe}{2}] emission line at 5.34\,µm anti-correlates strongly with $n$. The left panel of Figure~\ref{fig:FeIIvsH2vsn} presents the relationship between [\ion{Fe}{2}]/\hh\ S(1)--S(2) and $n$, which has a Spearman Rank coefficient equal to $-$0.72 with a $p$-value of 7.2~$\times$~10$^{-47}$. Iron has a high condensation temperature above 1000\,K and thus is typically found in the solid phase \citep{SS1996}. The strong anti-correlation of $n$ with [\ion{Fe}{2}] strength relative to lower-$J$ \hh\ emission is potentially stronger than that of other bright ionized lines due to enhancement of gas-phase iron abundance via shock heating in neutral gas, which can liberate it from its solid form \citep{Ohalloran2008}. Spatially, the [\ion{Fe}{2}] emission is correlated with but somewhat more extended than [\ion{Ne}{2}]+[\ion{Ne}{3}],
as seen in the contours of Figure~\ref{fig:allmaps}. 

The excitation of warmer molecular gas as traced by \hh\ S(4)--S(7) may have contributions from UV pumping \citep{BlackDalgarno1976}, increasing the strength of emission at low power-law index, $n$, compared to the collisionally excited lower $J$ transitions (S(1)--S(2)). Since the emission of both [\ion{Fe}{2}] and \hh\ S(4)--S(7) can be enhanced at low power-law index $n$, these two effects approximately cancel out, causing [\ion{Fe}{2}]/\hh\ S(4)--S(7) to be more weakly correlated with $n$. This relationship is shown in the right panel of Figure~\ref{fig:FeIIvsH2vsn}, with a Spearman Rank coefficient of $-$0.32 and a $p$-value of 4.9~$\times$~10$^{-8}$. Because the collisionally-dominated cooler phases of molecular gas are not enhanced at low $n$, [\ion{Fe}{2}] emission is stronger relative to \hh\ S(1)--S(2) than \hh\ S(4)--S(7), providing an indication of recent shock heating. 

Recent shock heating may also play a role in the suppression of the observed OPR at low $J$. The OPR thermalization timescale by \hh\,--H collisions proceeds at a rate proportional to the atomic H abundance. Shocks that partially dissociate \hh\ will therefore boost the thermalization rate. If the shocks are fast enough, high temperatures may suffice to raise the OPR to $\sim$3 via collisions with H atoms and protons. Slower shocks in cooler molecular gas (S(1)--S(2)) may increase the rotational temperature without dissociating \hh\, contributing to the OPR falling below its equilibrium value, as observed.

\section{Model Caveats} \label{sec:shortcomings}
\subsection{Model Over-estimations of S(6)}
\label{S6_overestimate}
The disequilibrium model (\S\ref{sec:ksimodel}) often mildly overestimates the column density of the S(6) transition. In Figure~\ref{fig:model_residuals}, it can be seen that the median residuals of model fits in the M82 center indicate the column density $N_{\mathrm{obs},\hh(J=6)}$ is overestimated by $\sim$15\%. We discuss several potential reasons for the model overestimation at S(6) below.

\subsubsection{6.02 µm H$_2$O Water Ice Absorption}
The S(6) transition occurs at a wavelength of 6.11\,µm. There is a H$_2$O ice absorption feature centered at 6.02\,µm with a FWHM of 0.58\,µm \citep{Gao2013} that has been detected in spectra of galaxies \citep{Spoon2002, Spoon2004, Lai2020}. The FWHM of this feature can vary due to the fact that it is composed of multiple overlapping components with different widths and carriers \citep{Boogert2015}. If the 6.02\,µm H$_2$O ice absorption feature is present, the FWHM of the 6.2\,µm PAH feature will be narrower than its empirical value. While we do detect the 6.2\,µm PAH feature with a narrower FWHM, we lack a rigorous process for determining the optical depth of the potential H$_2$O ice absorption.

\citet{Lai2020} investigates extinction from silicates and water in a large sample of nearby star-forming galaxies. They report median values for the optical depths of silicates and the 3.05\,µm H$_2$O ice feature in their sample. We convert this into a scaling between silicates and 6.02\,µm H$_2$O ice based on the band strengths provided by \citet{Gao2013}. This relation allows us to estimate the possible extinction by ice absorption near the S(6) transition. From this relationship, we find possible values of $\tau_\mathrm{{ice}}$ range from 6.3\,$\times$\,10$^{-7}$ to 0.19, with a median value of 0.07. Assuming screen geometry to correct for attenuation, we obtain a range of possible S(6) ``boost factors'' of 1.0--1.21, with a median value of 1.07. Since the actual strength of 6\,µm ice absorption is unknown, we add an additional source of uncertainty to the column density of S(6) appropriate for the median boost factor, resulting in an additional 7\% uncertainty, added in quadrature to $\mathrm{\sigma_{S(6)}}$. We then fit the observed column densities with the corrected $\mathrm{\sigma_{S(6)}}$ using our temperature-dependent $\xi$ model. We chose to boost $\mathrm{\sigma_{S(6)}}$ rather than correct the column density at S(6) because the MIRI/MRS data does not have coverage of the 3.05\,µm H$_2$O ice feature, so its strength cannot be determined definitively. Thus, the S(6) transition is weighted less when performing the fit to the disequilibrium model, indirectly reducing the median residuals at S(1) and S(4) from $<$ 15\% to $<$ 7\%. 

\subsubsection{Self-Shielding}
Steady-state photodissociation regions (PDRs) may also experience non-equilibrium OPRs with an OPR$>$3, where differences in self-shielding enhance the column density of the more dominant species \citep{DB1996,NMH1998, SN1999, LeBourlot1999, Hunt2025}. Due to its higher abundance (when OPR$>$1), ortho-\hh\ preferentially self-shields from photodestruction in low optical depth regions. This can result in the inferred column density of ortho-\hh\ appearing above the para-\hh\ levels on a SLED (implying $\xi>1$) as discussed in \citet{Hunt2025}. S(6) is a para-\hh\ transition, so its observed column density may be reduced in translucent regions with elevated FUV fluxes. \citet{Hunt2025} finds an OPR~$>$~OPR$\mathrm{_{eq}}$ in three photodissociation fronts in the nearby, low-metallicity dwarf galaxy, I~Zw~18, and attributes it to low \hh\ column densities and the preferential self-shielding of ortho-\hh. However, we do not detect any cases of OPR~$>$~OPR$\mathrm{_{eq}}$ ($\xi$ $>$ 1, equivalently) in the starburst region of M82, and thus we do not find supporting evidence for preferential self-shielding of ortho-\hh. 

\subsubsection{Thermalization Threshold Temperature}
\label{subsec:Teq}
In our temperature dependent disequilibrium model, we strictly enforce equilibration between the OPR and OPR$\mathrm{_{eq}}$ at gas temperatures $>$\,1000\,K (\S\ref{subsec:temp_dep_ksi_model}). This is an approximation based on the fact that the rate of para-to-ortho-\hh\ conversion has a strong temperature dependence which rises steeply at $\sim$\,1000\,K \citep{Neufeld2006}. Because the column density at S(6) is \textit{over}-estimated, it may hint that the OPR is thermalizing at temperatures below 1000\,K. This is supported by the fact that the models assuming OPR = OPR$\mathrm{_{eq}}$ in Figure~\ref{fig:model_residuals} (\citetalias{TS2016} and \citetalias{TS2016}+extinction) are the only models where the inner three quintiles of SLED fits accurately reproduce the observed column density at S(6). 

To test its impact, we lower the equilibration threshold temperature to $T_{\mathrm{eq}}$ = 500\,K in the temperature-dependent $\xi$ model (\S\ref{subsec:temp_dep_ksi_model}, Equation~\ref{eq:xi_Trot}). Median residuals at S(6) decrease from $\sim$28\% with $T_{\mathrm{eq}}$ = 1000\,K to $\sim$19\%. However, with four degrees of freedom, the overall fits are worse, with average reduced chi-square value increasing by $\sim$13\% for $T_{\mathrm{eq}}$ = 500\,K relative to $T_{\mathrm{eq}}$ = 1000\,K. This may indicate that OPR thermalization is proceeding over a range of gas temperatures 500\,K~$<$~$T_{\mathrm{eq}}$~$<$~1000\,K. 

With $T_{\mathrm{eq}}$ set to 500\,K, the median values of $n$ and \ts\ increase slightly by $\sim$~1.5\% and 5.4\% to 4.66\,$\pm$\,0.25 and 1.36\,$\pm$\,0.35, respectively. The median value of $\xi$ decreases by $\sim$~48\% to 0.33\,$\pm$\,0.10. This reduction in $\xi$ with $T_{\mathrm{eq}}$ = 500\,K compared to $T_{\mathrm{eq}}$ = 1000\,K (median $\xi$\,=\,0.49\,$\pm$\,0.06) reinforces the measured suppression in the OPR in the M82 nuclear starburst, but highlights the sensitivity of $\xi$ to the choice of $T_{\mathrm{eq}}$. 

Compared to the pure power-law plus extinction and single-$\xi$ models (for models with $T_{\mathrm{eq}}$ = 100\,K and $T_{\mathrm{eq}}$ = 2000\,K, respectively, \S~\ref{sec:modelval}), the average chi-square value increases by a factor of four and two, relative to $T_{\mathrm{eq}}$ = 1000\,K. Thus, $T_{\mathrm{eq}}$ = 1000\,K provides the best overall fit for OPR thermalization in the M82 nuclear starburst for the four values tested for $T_{\mathrm{eq}}$. Future models may seek to pin down more precisely the gas temperature dependence of OPR thermalization.

\section{Conclusions} \label{sec:conc}
In this work, we have built upon existing frameworks for modeling \hh\ excitation in galaxies using power-law temperature distributions (e.g., \citetalias{TS2016}) to incorporate dust extinction and the effects of non-uniform disequilibrium heating on the observed ortho-to-para ratios. Deviations from the equilibrium-value OPR reveal valuable information about the thermal history of the molecular gas, providing a time-average tracer of varying heating rates by identifying instances of rapid heating within galaxies. Our model accurately reproduces the distribution of observed column densities in the central regions of the nearby, edge-on, starburst galaxy, M82. The additional physical processes in our model permit its use in a wide range of environments, including recently heated and obscured gas, which previous models could not account for. With the rich MIR spectroscopy from JWST, a physically-motivated model of \hh\ excitation allows for a deeper understanding of the time-varying interactions occurring in the molecular gas. 

We find the following results:
\begin{itemize}
    \item We detect high signal-to-noise emission of pure rotational transitions of \hh\ from S(1) to S(7) in 413 regions in the MIRI/MRS coverage of the M82 nuclear starburst. The observed trends in column density show varying power-law slopes of their temperature distribution as well as departures from equilibrium-value OPRs indicated by a damping zig-zag shape in excitation diagrams. 

    \item Median residuals between our temperature-dependent disequilibrium model and observations are $<$ 18\% for all transitions S(1)--S(5) and S(7), reducing the median residuals relative to the pure power-law model by a factor of $\sim$5. Moderate remaining residuals at S(6) could result from complications such as the unknown presence of the 6.02\,µm water ice absorption feature, thermalization of the OPR at lower temperatures, and/or optically thin shielding effects at high $J$. 
    
    \item Attenuation of \hh\ rotational emission near 9.7\,µm (S(3)) due to silicate absorption is detected in all regions, with a median optical depth \ts$\sim$1.3\,$\pm$\,0.3. This value is inferred by the attenuation-driven deviations seen in the S(3) transition relative to the other detected transitions, producing attenuation estimates that closely track values inferred from the 5.5--14.5\,µm continuum with screen geometry.

    \item We find the mass of warm molecular gas ($T>100\,$K) to be (2.0\,$\pm$\,0.1)\,$\times$\,10$^{7}$\,M$_\odot$ across the JWST/MIRI MRS coverage of a $\sim$ 500~pc region of the M82 nuclear starburst, representing $\sim$7.6\% of the total molecular mass in the disk, as inferred from CO(1-0) observations.
    
    \item The median value for the power-law index of the temperature distribution, $n$, is found to be 4.58~$\pm$~0.20, consistent with but slightly warmer (lower $n$ by 5\%) than previous studies of galaxies within varying galaxy environments \citepalias{TS2016}.
    
    \item The OPR is found to be nearly uniformly about half of its equilibrium value across the starburst core.  This can be attributed to the turbulent environment of the M82 nuclear starburst where cool gas is heated and mixed on timescales shorter than the spin conversion timescale. 
    
    \item Relative to low-$J$ \hh\ line luminosity from S(1) and S(2), the strength of [\ion{Fe}{2}] emission is enhanced in gas where the temperature distribution is biased towards warmer gas (lower $n$). Only a weak enhancement is seen in ionized lines such as [\ion{Ne}{2}] and [\ion{Ne}{3}], which may suggest that shock heating is both liberating [\ion{Fe}{2}] from its solid phase while also increasing the temperature of the gas.

\end{itemize}

Future models that seek to understand the thermal history of gas within galaxies may find OPR disequilibrium provides a useful physical diagnostic for quantifying the effects of time varying heating conditions. Observationally, the connection between increased turbulence and OPR suppression may serve as a predictor for increased CO linewidths and low $\alpha_{CO}$.

\begin{acknowledgments}
This work is based on observations made with the NASA/ESA/CSA James Webb Space Telescope. We acknowledge support from program JWST-GO-01701 which is provided by NASA through a grant from the Space Telescope Science Institute, operated by the Association of Universities for Research in Astronomy, Inc., under NASA contract NAS 5-03127. The data were obtained from the Mikulski Archive for Space Telescopes at the Space Telescope Science Institute, operated by the Association of Universities for Research in Astronomy, Inc., under NASA contract NAS 5-03127 for JWST.
RSK and SCOG acknowledge financial support from the ERC via Synergy Grant ``ECOGAL'' (project ID 855130) and from the German Excellence Strategy via the Heidelberg Cluster ``STRUCTURES'' (EXC 2181 - 390900948). In addition RSK is grateful for funding from the German Ministry for Economic Affairs and Climate Action in project ``MAINN'' (funding ID 50OO2206), and from DFG and ANR for project ``STARCLUSTERS'' (funding ID KL 1358/22-1). V.V. acknowledges support from the Comité ESO Mixto 2024 and from the ANID BASAL project FB210003. R.H.-C. thanks the Max Planck Society for support under the Partner Group project "The Baryon Cycle in Galaxies" between the Max Planck for Extraterrestrial Physics and the Universidad de Concepción. R.H-C. also gratefully acknowledge financial support from ANID - MILENIO - NCN2024\_112 and ANID BASAL FB210003. This research has made use of NASA’s Astrophysics
Data System Bibliographic Services.

\facilities{JWST (MIRI)}

\software{Astropy \citep{astropy:2013, astropy:2018, astropy:2022}, MatPlotLib \citep{matplotlib}, NumPy \citep{numpy}, and SciPy \citep{scipy}}.
\end{acknowledgments}

\vspace{5mm}

\bibliography{refs}{}
\bibliographystyle{aasjournalv7}

\end{document}